\documentclass[twoside]{article}

\usepackage[sc]{mathpazo} 
\usepackage[T1]{fontenc} 
\usepackage{microtype} 

\usepackage[hmarginratio=1:1,top=32mm,columnsep=20pt,left=2.5cm]{geometry} 
\usepackage{multicol} 
\usepackage[hang, small,labelfont=bf,up,textfont=it,up]{caption} 
\usepackage{booktabs} 
\usepackage{float} 
\usepackage[colorlinks=true,linkcolor=black,citecolor=black,urlcolor=black]{hyperref} 

\usepackage{lettrine} 
\usepackage{paralist} 

\usepackage{abstract} 

\usepackage{titlesec} 
\renewcommand\thesection{\Roman{section}}
\titleformat{\section}[block]{\large\scshape\centering}{\thesection.}{1em}{} 

\usepackage{fancyhdr} 
\usepackage[]{natbib}
\usepackage{graphicx}

\title{\vspace{-15mm}\fontsize{24pt}{10pt}\selectfont\textbf{Extracting scaling laws from numerical dynamo models}} 

\author{
\large
\textsc{Z. Stelzer and A. Jackson}\\[2mm] 
\normalsize Earth and Planetary Magnetism Group, ETH Zurich, Switzerland\\ 
\normalsize \href{mailto:zacharias.stelzer@erdw.ethz.ch}{zacharias.stelzer@erdw.ethz.ch} 
\vspace{-5mm}
}
\date{}

\begin{document}

\maketitle 

\thispagestyle{fancy} 


\begin{abstract}

\noindent Earth's magnetic field is generated by processes in the electrically conducting, liquid outer core, subsumed under the term `geodynamo'. In the last decades, great effort has been put into the numerical simulation of core dynamics following from the magnetohydrodynamic (MHD) equations. However, the numerical simulations are far from Earth's core in terms of several control parameters. Different scaling analyses found simple scaling laws for quantities like heat transport, flow velocity, magnetic field strength and magnetic dissipation time.\\
We use an extensive dataset of 116 numerical dynamo models compiled by Christensen and co-workers to analyse these scalings from a rigorous model selection point of view. Our method of choice is leave-one-out cross-validation which rates models according to their predictive abilities. In contrast to earlier results, we find that diffusive processes are not negligible for the flow velocity and magnetic field strength in the numerical dynamos. Also the scaling of the magnetic dissipation time turns out to be more complex than previously suggested. Assuming that the processes relevant in the numerical models are the same as in Earth's core, we use this scaling to estimate an Ohmic dissipation of 3-8~TW for the core. This appears to be consistent with recent high CMB heat flux scenarios.

\end{abstract}


\begin{multicols}{2} 

\section{Introduction}
\label{sec:introduction}

The Earth's magnetic field is generated by motions of an electrically conducting fluid in the outer core, the bulk being liquid iron. The processes include magnetic induction and are subsumed under the term `geodynamo'. It is generally accepted that the fluid motions in the outer core, which are most important for maintaining the geodynamo, are driven by convection, i.e. by thermal and compositional buoyancy forces \citep{Olson:2007p17}. There are in general three ways to study the dynamics of the outer core. The first builds on theoretical considerations like force balances and thermodynamics \citep[e.g.][]{Jones:2011p4}. Secondly, it is possible to model the whole system numerically on the basis of the fundamental physical equations. Finally, laboratory experiments analogue to the processes proposed for the Earth's core can help to determine certain aspects of the dynamics. In this paper, we focus on the second approach.\\
An important part of the increase in knowledge about core dynamics in the last two decades came from numerical simulations of the dynamo process. Starting from the first successful 3D MHD self-sustained dynamo models of \citet{Glatzmaier:1995p2071} and \citet{Kageyama:1995p2949}, numerical dynamo simulations have been able to reproduce various features of the geo\-mag\-netic field such as field morphology, secular variations and polarity reversals. The problem, however, remains how to apply results from numerical simulations to the Earth.\\
A major challenge is the discrepancy between numerical models and the core in terms of the non-dimensional parameters defined in Table \ref{tab:nondim_no}. Specifically, numerical dynamos have far too slow rotation (Ekman number too large), are less turbulent (Rayleigh number too small) and excessively viscous relative to their electrical conductivity (magnetic Prandtl number too large) compared to the core. This gap can not be bridged easily due to the enormous computational power required to resolve all relevant time and length scales.\\
One way of using present-day numerical dynamo simulations to estimate quantities that are relevant to Earth's core (e.g. heat flux, flow velocity, magnetic field strength) is to extract scaling laws between these quantities and other characteristic parameters from the data. Assuming that the relevant processes in the core are the same as in our simulations, we may extrapolate the results to the parameter regime of the core and in that way gain insight into the processes in Earth's core.\\
This has been done for various quantities. Important results were the diffusivity-free scalings of heat transport, flow velocity and magnetic field strength \citep{Christensen:2006p11} and simple scalings for the magnetic dissipation time \citep{Christensen:2004p706,Christensen:2010p1}. The question arises, however, how complex a model needs to be in order to do justice to the data. \\
We address the classical problem of model selection, where a model is defined in terms of a number of parameters. On the one hand, the paradigm of Occam prefers a model that is less complex over another that is more complex (when both fit the data equally well), generally meaning that the former model contains the fewest parameters of all models. But what is often not recognised, and is equally important, is that models with fewer parameters can have greater predictive power than more complicated models. Physical theories are not only validated by their fit to existing data, but even more by their performance in predicting new data. A few words are in order to motivate why this phenomenon is true.\\
We imagine a noisy dataset with $n$ points and fit it with $p$ parameters; we begin by taking $p=n$ to achieve a perfect fit to our data. Because of noise, this model is extremely complex, containing high frequency oscillations (in the case of a function $f(x)$ fitted to points distributed in $x$). Imagine now receiving a new datum. The $n$ parameter model will have almost no predictive power for this new datum, since it has fitted all of the noise in the dataset from which it was derived. Indeed, a far simpler model, with $p\ll n$ will have far greater predictive power. We use this principle by implementing a procedure called `leave-one-out cross validation', where we systematically omit one of the data points and hold it in reserve as a test point, against which different models can test their predictive power. In this way we evaluate the predictive power of models, and find models based on an optimal number of parameters that have the most predictive power.\\
The format of the paper is as follows: In section \ref{sec:dynamo_dataset}, we present the database used in our analysis. In section \ref{sec:cross_validation}, we illustrate the method of cross-validation with a toy problem, before going on to apply it to the dynamo problem at hand. Subsequently, we analyse the scaling laws for heat transport, flow velocity and magnetic field strength using diffusivity-free parameters (section \ref{sec:diff_free_scal}) and traditional non-dimensional numbers (section \ref{sec:trad_scal}). Section \ref{sec:mag_dissipation_core} is concerned with the scaling of magnetic dissipation time as well as the application of the scalings to the core.

\section{Dynamo dataset}
\label{sec:dynamo_dataset}

\subsection{Numerical dynamo simulations}
\label{sec:num_dynamo_sim}

In the numerical dynamo simulations used in this study, convection is driven by a fixed superadiabatic temperature contrast $\Delta T$ between inner and outer boundaries of a rotating spherical shell. Moreover the Boussinesq approximation is used, i.e. density variations enter the equations only through a buoyancy term in the momentum equation.
The standard set of equations consists of five equations describing conservation of momentum (Navier-Stokes equation), magnetic induction, the transport of temperature and the solenoidal nature of the magnetic field $\mathbf{B}$ and the velocity field $\mathbf{u}$ (cf. eq. \ref{eqn:NS_eq}-\ref{eqn:solenoidal_u}).\\ 
These equations can be non-dimensionalised by introducing four independent control parameters. Their choice is not unique. We follow \citet{Christensen:2006p11} and use the shell thickness $D=r_o-r_i$ of the outer core, the inverse rotation rate $\Omega^{-1}$ , the temperature difference $\Delta T$, and the quantity $(\rho \mu_0)^{1/2} \Omega D$ as fundamental scales for length, time, temperature and magnetic field, respectively; $r_o$ is the outer core radius, $r_i$ the inner core radius, $\rho$ density and $\mu_0$ magnetic permeability. This leads to the following set of non-dimensional equations for magnetic field $\mathbf{B}$, fluid velocity $\mathbf{u}$ and temperature $T$:
\begin{eqnarray}
\label{eqn:NS_eq}
\frac{\partial \textbf{u}}{\partial t} + (\textbf{u} \cdot \nabla) \textbf{u} + 2(\hat{\textbf{z}} \times \textbf{u}) + \nabla \Pi =  \nonumber\\ 
\hspace{0.3cm} Ra \hspace{1pt} Ek^2 \hspace{1pt} Pr^{-1} \frac{\textbf{r}}{r_o}T + (\nabla \times \textbf{B}) \times \textbf{B} + Ek \hspace{1pt} \nabla^2 \textbf{u}	
\end{eqnarray}
\begin{eqnarray}
\label{eqn:mag_ind_eq}
\frac{\partial \textbf{B}}{\partial t} &=&\nabla \times (\textbf{u} \times \textbf{B}) + Ek \hspace{1pt} Pm^{-1} \hspace{1pt} \nabla^2 \textbf{B}
\end{eqnarray}
\begin{eqnarray}
\label{eqn:temp_transport_eq}
\frac{\partial T}{\partial t} + (\textbf{u} \cdot \nabla) T &=& Ek \hspace{1pt} Pr^{-1} \hspace{1pt} \nabla^2 T
\end{eqnarray}
\begin{eqnarray}
\label{eqn:solenoidal_B}
\nabla \cdot \textbf{B} &=& 0
\end{eqnarray}
\begin{eqnarray}
\label{eqn:solenoidal_u}
\nabla \cdot \textbf{u} &=& 0,
\end{eqnarray}
where $\hat{\textbf{z}}$ is the unit vector in the direction of the rotation axis. In these equations, gravity is assumed to vary proportional to the radius, $g_o$ being the value of gravity at the outer boundary; volumetric heating is neglected and $\Pi$ is the non-hydrostatic pressure. The four non-dimensional parameters governing equations \ref{eqn:NS_eq}-\ref{eqn:solenoidal_u} are defined in Table \ref{tab:nondim_no}.
\begin{table*}
\begin{minipage}{0.9\textwidth}
\caption{Non-dimensional parameters, their estimated values for Earth's core \citep[following][]{Olson:2007p17} and values in the models studied here. The first four quantities are input parameters to the numerical simulations, the lower ones are output parameters. \newline
$U$ is a characteristic velocity; $\nu$ is kinematic viscosity; $k$ is thermal conductivity; $\kappa$ is thermal diffusivity; $\alpha$ is thermal expansivity; $\eta = (\mu_0 \sigma)^{-1}$ is magnetic diffusivity with $\sigma$, electrical conductivity; $Q$ is heat flux; the remaining quantities are defined in the text. \newline
Note that the thermal diffusivity $\kappa$ and the electrical conductivity $\sigma$ have recently been revised. These ab-initio calculations have increased the numerical values of $\kappa$ and $\sigma$ by roughly a factor of three \citep{Pozzo:2012p2852,deKoker:2012p2910}. As a result, the non-dimensional parameters depending on those quantities have been revised with respect to those given in \citet{Olson:2007p17}. We give the updated numbers for $Pr$, $Pm$ and $Rm$.}
\label{tab:nondim_no}
\centering
\begin{tabular}{c c c c}
\hline
Quantity 			& Definition							& Earth's core 				& This study \\ \hline
Ekman 			& $Ek = \nu / \Omega D^2$				& $\sim 3\cdot10^{-14}$		& $10^{-6} - 10^{-3}$ \\
Rayleigh			& $Ra = \alpha g_o \Delta T D^3 / \nu \kappa$	& $\sim 10^{20\pm?}$		& $3\cdot 10^5 - 2.2\cdot 10^9$ \\
Prandtl			& $Pr = \nu / \kappa	$					& $\sim 0.1$				& $0.1 - 10$ \\
Magnetic Prandtl	& $Pm = \nu / \eta $						& $\sim 3\cdot 10^{-5}$		& $0.06 - 33.3$ \\ \hline
Nusselt 			& $Nu = Q D / 4\pi r_o r_i k \Delta T$		& ?						& $2.02 - 29.8$ \\
Magnetic Reynolds	& $Rm = U D / \eta$						& $\sim 2300$				& $39 - 5695$ \\ \hline
\end{tabular}
\end{minipage}
\end{table*}%
\\
For our analysis of scaling laws, we use a database of 185 numerical dynamo models built over time by U. Christensen and co-workers. Most of the models were previously reported in \citet{Christensen:2006p11} and \citet{Christensen:2009p704}, and studied in \citet{Christensen:2010p1} and \citet{King:2010p5}. The mechanical boundary conditions are no-slip and the ratio between inner and outer core radius is 0.35 as in Earth's core. The inner core of the models is insulating in some simulations and conducting in others. The exterior of the shell is electrically insulating in all cases. We restrict our analysis to this database, which is homogeneous in terms of model setup and numerical method, in order to avoid unwanted effects of varying too many control parameters in the scaling law selection.

\subsection{Scaling laws and model setup}
\label{sec:scal_laws_model_setup}

We seek to extract scaling laws from numerical solutions of the MHD equations \ref{eqn:NS_eq}-\ref{eqn:solenoidal_u} as explained in the introduction. Under certain conditions, these scaling laws may then be extrapolated to the more extreme parameter range of Earth's core. An example of a scaling law is the classical heat transport ($Nu$-$Ra$) scaling in non-rotating, plane-layer convection. The functional relationship between $Nu$ and $Ra$ can be expressed as $Nu \sim Ra^\beta$ with possibly different values of $\beta$ for different convective regimes \citep[e.g.][]{Aurnou:2007p1362}.\\
Similarly, we follow the ground-breaking work of \citet{Christensen:2006p11} and others and restrict our scaling analysis to power laws of the from
\begin{equation}
\label{eqn:exp_scal_form}
	\hat{y} = \alpha \hspace{2pt} \prod_{j=1}^{p-1} x_j^{\beta_j}.
\end{equation}
Observations are collected in $y$ and are the output of the numerical simulations; predictions $\hat{y}$ in equation \ref{eqn:exp_scal_form} are calculated from $x_j$, the independent variables, which are mostly control parameters of the MHD equations. The number of data (numerical dynamo simulations) and thereby the size of $\hat{y}$ is $n$; the total number of free parameters is $p$ consisting of the prefactor $\alpha$ and $(p-1)$ exponents $\beta_j$.\\
The task of fitting this functional form to given data can be transformed to a linear problem by taking the logarithm,
\begin{equation}
\label{eqn:log_scal_form}
	\log \hat{y} = \log \alpha +  \sum_{j=1}^{p-1} \beta_j \log x_j.
\end{equation}
Our linear model includes the coefficients $\log \alpha$ and $\beta_j$. These are fitted by multiple linear regression which minimizes the mean quadratic misfit,
\begin{equation}
\label{eqn:chi_squared}
	\chi^2 = \frac{1}{n} \sum_{i=1}^n \left(\frac{\zeta_i - \hat{\zeta}_i}{\sigma_i} \right)^2,
\end{equation}
where we have defined $\zeta=\log y$ for ease of notation. The contribution of the different data points to $\chi^2$ can be weighted by their standard deviation $\sigma_i$.\\
As another measure of misfit between data and fitted values, we define the mean relative misfit to the original data $y$ (not in $\log$-domain),
\begin{equation}
	\label{eqn:mean_rel_misfit}
	\chi_{rel} = \sqrt{\frac{1}{n} \sum_{i=1}^n \left( \frac{y_i - \hat{y}_i}{y_i} \right) ^2},
\end{equation}
for comparability with \citet{Christensen:2006p11}.

\subsection{Errors in the dependent variable}
\label{sec:errors}

We seek to fit the linear model to observed values $\zeta$, but, in doing so, we face the question of what the appropriate attribution of errors for these observations is. In principle the numerical experiments are perfect, and it may be our parametrised theory that is an imperfect representation of the data. Obvious error sources are the limited resolution of the models and the limited time averaging of fluctuating properties; but equally there may be errors in the observations as a result of the simulations perhaps not achieving equilibrium, or perhaps as a result of bistability and/or hysteresis in the nonlinear system \citep[see, for example,][]{Simitev:2009p2995}. Two routes are available to us: following \citet{Christensen:2006p11}, we can assume that the errors are equal in $\zeta = \log y$, or we could alternatively assume that the errors are equal in the original measured variable $y$. The first hypothesis leads to the error $\sigma_\zeta=c$, where $c$ is constant; one can see, from a consideration of the perturbation $\delta(\log y)$, that this leads to $\delta y/y=c$, namely that the {\sl percentage} errors in the original observations $y$ are constant. Whether this is a good model remains open. The second assumption, that there are constant errors $\sigma_y$ in the original observations $y$, leads to
\begin{equation}
	\sigma_\zeta=\sigma_y/y=\sigma_y/e^\zeta,
\end{equation}
when the errors are small. In this model, the errors shrink drastically when $\zeta$ is large. In the absence of definitive knowledge concerning the errors, we choose to carry out fitting using both attributions of error. In the following sections, we assume equal errors in $\zeta$. The results under the assumption of equal errors in the original variable $y$ are given in Appendix \ref{sec:equal_errors_y}. Considering the resulting error distributions, it is still not clear which error attribution is appropriate.

\subsection{Parameter range}
\label{sec:par_range}

\begin{figure*}
\centering
\includegraphics[width=\columnwidth]{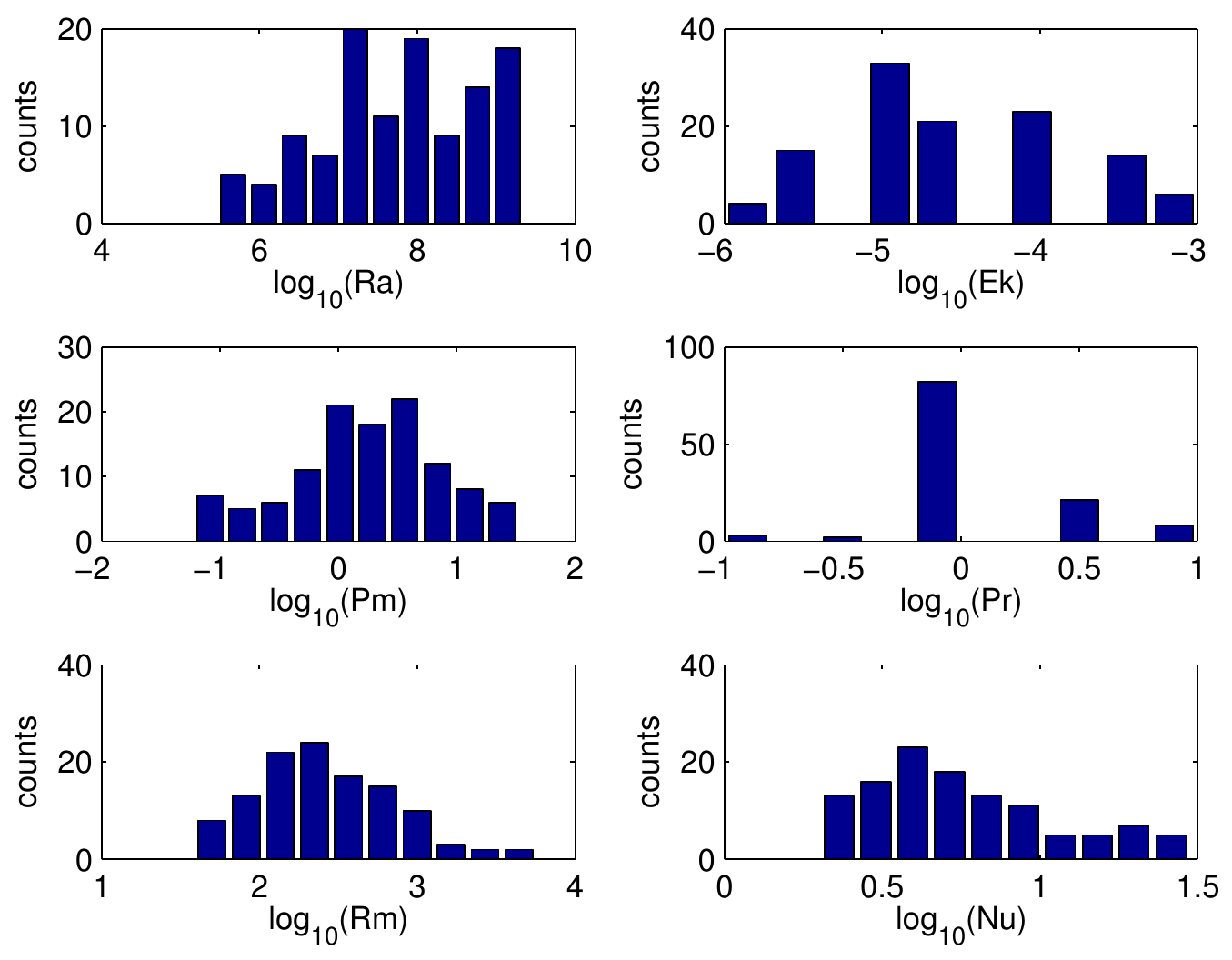}
\caption{Histograms of the values of the non-dimensional parameters in the 116 simulations used in the scaling law analysis. All parameters apart from $Pr$ show a distribution that allows the extraction of scaling laws.}
\label{fig:histograms}
\end{figure*}

For the extraction of scaling laws from the dynamo database, we only use simulations that satisfy the following criteria \citep[following][]{Christensen:2006p11}:
\begin{enumerate}
 \item The simulation must be fully convective as required by $Nu > 2$.
 \item The generated magnetic field has to be dipole-dominated. As a measure of dipolarity, we use $f_{dip} = B_{dip}/B_{12}$, the time-averaged ratio of the mean dipole field strength to the field strength in harmonic degrees 1 to 12 on the outer boundary. The condition for a dipole-dominated field is taken as $f_{dip} > 0.35$. 
 \item The Prandtl number should not fall too far from the values estimated for Earth's core: $Pr \le 10$. (Models in the dataset with $Pr > 10$ are rather new and have not been used by any other study.)
\end{enumerate}
Applying these restrictions to the data, we are left with 116 numerical dynamo simulations. We also tested excluding the models with the highest Ekman numbers, $Ek = 10^{-3}$, as done in \citet{Christensen:2006p11}. However, this hardly changed the result of our analysis. In section \ref{sec:dynamical_regime}, we will determine the effect of the requirements on $Nu$, $f_{dip}$ and $Pr$.\\
The 116 numerical dynamo simulations contain 40 models with an imposed two- or four-fold symmetry. We tested the effect of discarding those and found the same scaling laws as for the full dataset (section \ref{sec:diff_free_scal}), with the exponents just slightly changed.\\
Figure \ref{fig:histograms} shows the distribution of the control parameters $Ra$, $Ek$, $Pm$ and $Pr$ as well as the output quantities $Rm$ and $Nu$ within the 116 models used in the scaling law analysis. In general, the distribution of the parameter values appears to be sufficiently uniform over some range to allow the extraction of scaling laws. Only in the case of $Pr$, the values cluster at $Pr=1$ with very few differing values. Hence the database is not favourable to elicit a $Pr$-dependence. If we really were to apply the scaling laws to the Earth, $Pr$ fortunately is the number that requires the least extrapolation (cf. Table \ref{tab:nondim_no}).

\subsection{Dynamical regime}
\label{sec:dynamical_regime}

\begin{figure*}
\centering
\includegraphics[width=\columnwidth]{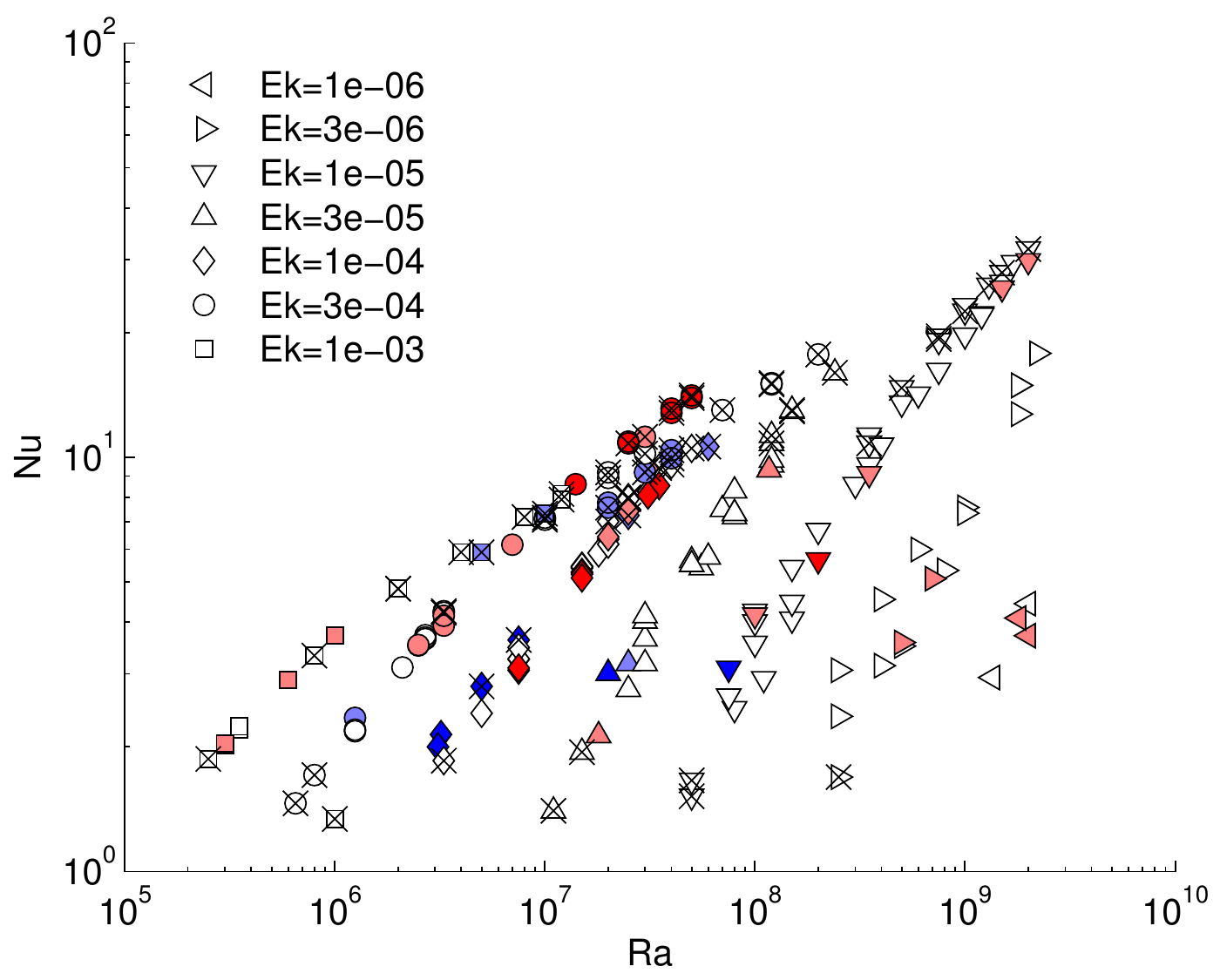}
\caption{Plot of $Nu$ vs. $Ra$ for all 185 dynamo models of the database. Colour indicates the value of $Pr$: dark-blue $Pr\le0.1$, light-blue $0.1<Pr<1$, white $Pr=1$, light-red $1<Pr<10$, dark-red $Pr\ge10$. Crossed-out models do not fulfill the criteria of section \ref{sec:par_range}. Note that the remaining 116 models fall into the rapidly-rotating regime.}
\label{fig:C185_Nu_Ra_crossed_colPr}
\end{figure*}
\begin{figure*}
\centering
\includegraphics[width=\columnwidth]{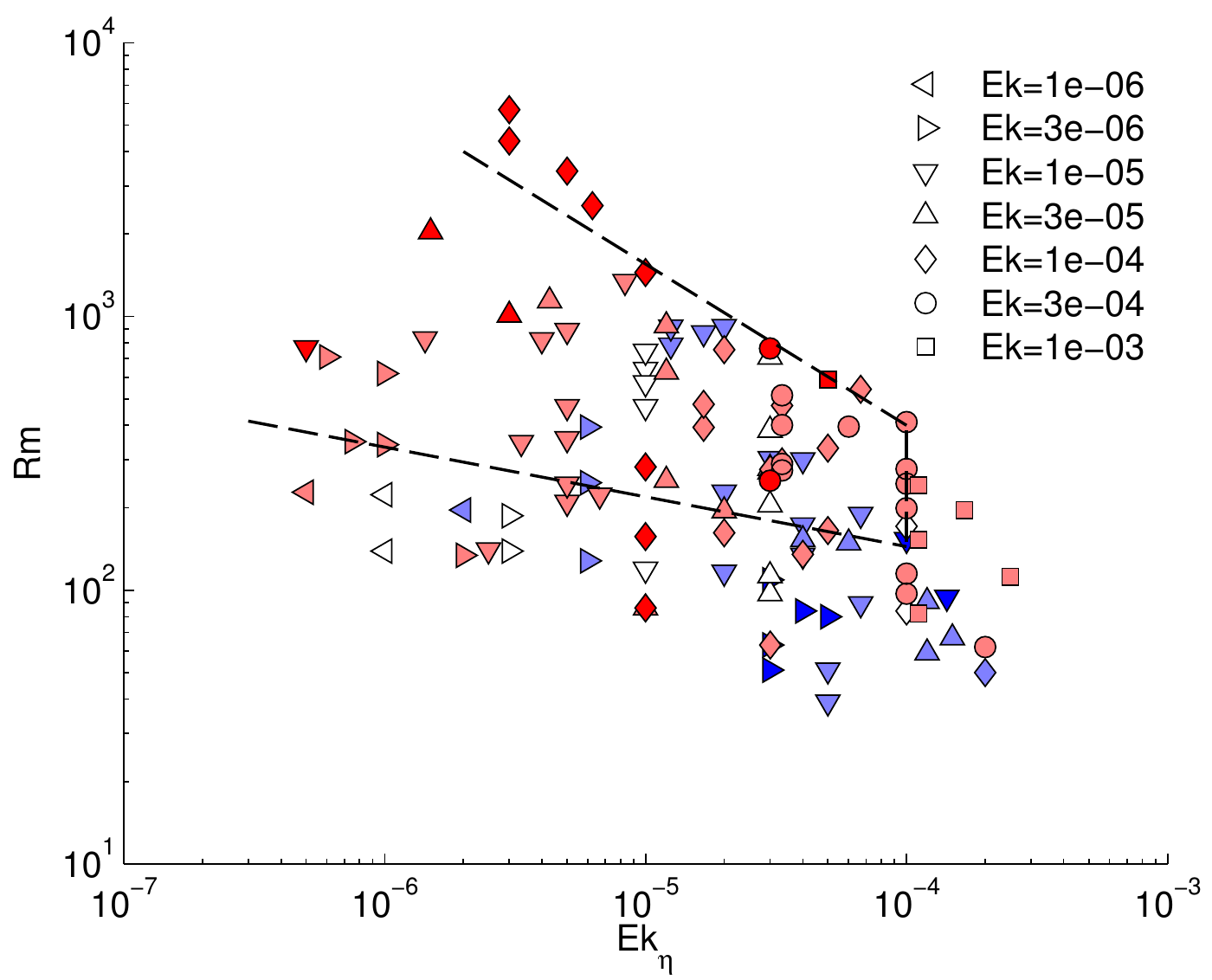}
\caption{Earth-likeness of the 116 dynamo models used in this study according to the criteria of \citet{Christensen:2010p2784}. $Ek_\eta = Ek/Pm$ is the magnetic Ekman number. Models that exhibit an Earth-like magnetic field morphology plot inside the dashed line. Colour indicates the value of $Pm$: dark-blue $Pm\le0.1$, light-blue $0.1<Pm<1$, white $Pm=1$, light-red $1<Pm<10$, dark-red $Pm\ge10$.}
\label{fig:C116_Rm_Eketa_triangle}
\end{figure*}

Convective heat transfer can be separated into two regimes, the rapidly-rotating and the buoyancy-dominated regime \citep{Aurnou:2007p1362}. In the rapidly-rotating regime, the flow is largely two-dimensionalised by the Taylor-Proudman theorem. For stronger forcing, buoyancy breaks the columnar structure leading to three-dimensional convective structures \citep{King:2009p739}. The two regimes are characterized by different heat transport efficiencies and different slopes in a plot of $Nu$ versus $Ra$.\\
Figure \ref{fig:C185_Nu_Ra_crossed_colPr} shows the quantities $Nu$ versus $Ra$ for the models in our database. Crossed-out models are rejected by the criteria in section \ref{sec:par_range}. Obviously, the majority of the 185 dynamo models falls into the rapidly-rotating regime. By applying the criteria on $f_{dip}$ and $Pr$, we throw out the models that are slightly buoyancy-dominated or transitional. The criterion $Nu>2$ would appear not to make a great difference were it not applied. As a result, we are left with 116 rapidly-rotating models for our analysis.\\
There have been attempts to classify geodynamo models according to their Earth-likeness. \citet{Christensen:2010p2784} used four criteria based on magnetic field morphology, namely relative axial dipole power, equatorial symmetry, zonality and flux concentration. They found that Earth-like dynamo models fall into a certain area in the ($Rm$-$Ek_\eta$) domain, where $Ek_\eta = Ek/Pm$ is the magnetic Ekman number. Figure \ref{fig:C116_Rm_Eketa_triangle} shows where the 116 dynamo models of this study plot in terms of $Ek_\eta$ and $Rm$. According to the criteria of \citet{Christensen:2010p2784}, 61 of the models have a magnetic field morphology that is Earth-like. We applied our scaling law analysis also to this subset of the data. The resulting scaling laws are given in Appendix \ref{sec:earth_like_models}. They are very similar to the ones in section \ref{sec:diff_free_scal} using all 116 dynamo models.


\section{Cross-validation}
\label{sec:cross_validation}

\subsection{Model selection}

Extracting scaling laws from multivariate data is a model selection problem, or more specifically, a variable subset selection problem. In section \ref{sec:scal_laws_model_setup}, we have defined the functional form of the scalings of interest (eq. \ref{eqn:exp_scal_form}). The question now is which independent variables $x_j$, should be included in the linear model (eq. \ref{eqn:log_scal_form}) in order to explain the values of the dependent variable $\hat{y}$.\\
The solution to this problem is not trivial. Normally one wishes to examine the discrepancy between theory and observation through a quantity such as mean quadratic misfit $\chi^2$ (eq. \ref{eqn:chi_squared}). In a linear problem, however, it is always possible to reach $\chi^2=0$ with $p \le n$, the number of free parameters less or equal to the number of data. Model selection ideally avoids over-fitting, so the model contains `just the right' (number of) independent variables in the sense that the model accounts for the variability in the data but is not more complex than required (Occam's razor). In the introduction we explained how it is possible for simpler models to have more predictive power than complex ones, and this is the property we seek to exploit.\\
A variety of approaches exists in the areas of frequentist and Bayesian statistics to tackle the task of model selection. An elegant way of determining the required independent variables $x_j$ for a model is cross-validation (CV). It is probably the simplest method for estimating prediction error \citep{Hastie:2009}. High predictive power, in turn, is certainly a desirable property for a scaling law.\\

\subsection{Leave-one-out cross-validation}

We use leave-one-out cross-validation (LOOCV) in our analysis. One observation of the $n$ data is set aside as a validation sample. The parameters of the linear model, $\log \alpha$ and $\beta_j$, are estimated (`trained') from the remaining $(n-1)$ data (training sample) by minimizing mean quadratic misfit $\chi^2$ (eq. \ref{eqn:chi_squared}). Then the model is validated by applying it to the validation sample. This process is done consecutively, setting aside a different part of the data and predicting it from the remainder. The misfit between the validation data point and its prediction from the corresponding model is accumulated, leading to the cross-validation estimate of the prediction error,
\begin{equation}
\label{eqn:P_CV}
	P_{CV} = \frac{1}{n} \sum_{i=1}^n \left( \frac{y_i - \hat{y}_i^*}{\sigma_i} \right)^2,
\end{equation}
where the prediction $\hat{y}_i^*$ has been obtained using the model that was trained on all but the $i$-th datum. The CV estimate of prediction error, $P_{CV}$, is calculated for models containing different combinations of independent variables, $x_j$. The favoured variable combination is the one with minimum $P_{CV}$. The parameters of the final scaling law are trained on all $n$ data.\\
Various other model selection methods such as Akaike's information criterion (AIC), Mallows' $C_p$, the jackknife and the bootstrap, are asymptotically equivalent to LOOCV \citep{Stone:1977,Efron:1983p2748}. A generalization of LOOCV is $k$-fold CV with $k$ instead of $n$ partitions. We experimented with different $k$. For the main purpose of this paper, however, the resulting differences are minor.

\subsection{Example: Curve fitting}

\begin{figure*}
\centering
\includegraphics[width=\columnwidth]{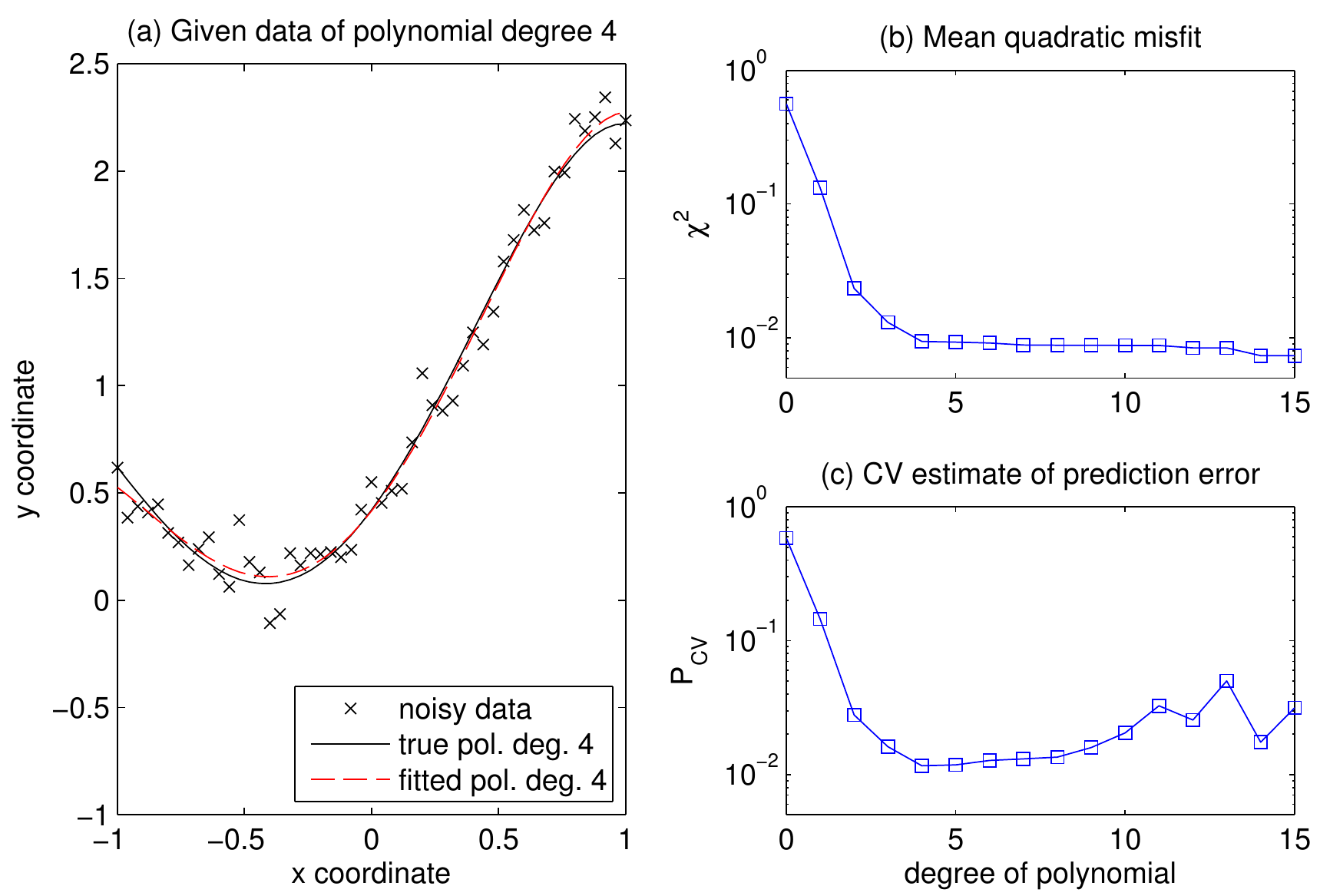}
\caption{Curve fitting, synthetic example. (a) Black crosses are 51 noisy data that were created from a Chebyshev polynomial of degree $m=4$ (black line) by adding Gaussian noise with standard deviation $\sigma=0.1$. The dashed red curve is the final fitted polynomial. (b) Mean quadratic misfit $\chi^2$ for polynomials of degrees 0 to 15. (c) LOOCV estimate of prediction error $P_{CV}$ for the same polynomial degrees. $\chi^2$ is successively reduced by increasing polynomial degree, whereas $P_{CV}$ is minimum for the true polynomial degree. For numerical values see Table \ref{tab:CV_pol_example}.}
\label{fig:CV_pol_deg4}
\end{figure*}

In order to illustrate the problem of model selection and how it can be solved by LOOCV, we give a synthetic example from the domain of curve fitting, which in this case also is a linear problem. Let us suppose we are given noisy data $y$ and all we know is that the data come from a model in the form of a Chebyshev expansion 
\begin{equation}
	y = \sum_{i=0}^m \beta_i T_i(x) + \epsilon,
\end{equation}
where $T_i$ are Chebyshev polynomials and $\epsilon$ is the noise. Now, we want to retrieve the underlying functional form and especially determine the degree $m$ of the underlying polynomial.\\
Figure \ref{fig:CV_pol_deg4}(a) shows 51 noisy data points that were created from a Chebyshev polynomial of degree $m=4$ by adding Gaussian noise with standard deviation $\sigma_{true} = 0.1$. The polynomial coefficients are listed in Table \ref{tab:CV_pol_example_coeff}. As in the applications later in this study, the amplitudes of the contributions from different polynomial degrees differ significantly. \\
Figure \ref{fig:CV_pol_deg4}(b) gives the mean quadratic misfit $\chi^2$ (eq. \ref{eqn:chi_squared}), assuming $\sigma=1$ out of ignorance, for multiple linear regressions using polynomials of degrees 0 to 15; the corresponding numerical values are given in Table \ref{tab:CV_pol_example}. The misfit $\chi^2$ can, of course, be reduced successively by using polynomials of higher degrees and falls to 0 for a polynomial of degree 50, when $p=n$, the number of free parameters $p$ equals the number of unknowns $n$.\\
Figure \ref{fig:CV_pol_deg4}(c) shows the LOOCV estimate of prediction error $P_{CV}$ (eq. \ref{eqn:P_CV}) for polynomials of degrees 0 to 15. The corresponding numerical values in Table \ref{tab:CV_pol_example} show that minimum $P_{CV}$ is reached for polynomial degree 4. LOOCV is also able to correctly identify the noise in the data. For the correct degree 4 polynomial, the noise level is found to be $\sqrt{\chi^2} \approx 0.097$ (cf. Table \ref{tab:CV_pol_example}). This value is better than for any other polynomial degree, the true answer being $\sigma_{true} = 0.1$. \\
The model selection procedure by LOOCV chooses the right degree $m$ of polynomial by rating the different models according to their predictive abilities. Moreover, the subsequently estimated polynomial coefficients $\hat{\beta}$ and the estimated noise level are quite close to their true values $\beta_{true}$ and $\sigma_{true}$, respectively.
\begin{table*}
\caption{Curve fitting, synthetic example. True polynomial coefficients $\beta_{true}$ used in the synthetic example, and their multiple linear regression estimates $\hat{\beta}$.}
\centering
\begin{tabular}{|c|c c|}
\hline
Pol. degree		& $\beta_{true}$ 	& $\hat{\beta}$ \\ \hline
0				& 1				& 1.0000 \\
1   				& 1				& 1.0335 \\
2   				& 0.5			& 0.4921 \\
3 				& -0.2			& -0.1590 \\
4 			 	& -0.08			& -0.0922 \\
\hline
\end{tabular}
\label{tab:CV_pol_example_coeff}
\end{table*}%
\begin{table*}
\caption{Curve fitting, synthetic example. Values of mean quadratic misfit $\chi^2$ and LOOCV estimate of prediction error $P_{CV}$ for polynomials of degrees $m$ from 0 to 8 and 48 to 50, see also Figures \ref{fig:CV_pol_deg4}(b) and \ref{fig:CV_pol_deg4}(c). At polynomial degree $m=50$, the number of free parameters equals the number of data, $p=n=51$. Minimum values are bold.}
\centering
\begin{tabular}{|c|c c|}
\hline
Pol. degree		& $\chi^2$		& $P_{CV}$ 	 \\ \hline
0				& 0.5624			& 0.5851 		 \\
1   				& 0.1324			& 0.1454 		 \\
2   				& 0.0234			& 0.0279 		 \\
3 				& 0.0131			& 0.0161 		 \\
4 				& 0.0094			& \bf{0.0116} 	 \\
5 				& 0.0093			& 0.0118 		 \\
6 				& 0.0091			& 0.0127 		 \\
7 				& 0.0088			& 0.0131 		 \\
8   				& 0.0088			& 0.0135 		 \\
\vdots 			& \vdots			& \vdots		 \\
48				& 3.4620e-05		& 4.5578e+22	 \\
49				& 1.1865e-05		& 1.1348e+24	 \\
50				& \bf{0}			& -		 	 \\
\hline
\end{tabular}
\label{tab:CV_pol_example}
\end{table*}%


\section{Diffusivity-free scalings}
\label{sec:diff_free_scal}

Following \citet{Christensen:2002p1341}, there have been several studies advocating diffusivity-free scaling laws for the application to Earth's core \citep{Christensen:2006p11,Christensen:2009p704,Christensen:2010p1}. The underlying physical rationale is the hypothesis that diffusive processes do not play a primary role in Earth's core. Hence a modified Nusselt number,
\begin{eqnarray}
	Nu^* &=& \frac{1}{4\pi r_o r_i} \frac{Q_{adv}}{\rho c \Delta T \Omega D} \nonumber \\
	&=& (Nu-1) \frac{Ek}{Pr},
	\label{eqn:Nustar}
\end{eqnarray}
has been introduced, where $Q_{adv}$ is advected heat flux and $c$ is heat capacity; the remaining quantities were defined in section \ref{sec:num_dynamo_sim}. Moreover a modified Rayleigh number,
\begin{eqnarray}
	Ra^* &=& \frac{Ra \hspace{2pt} Ek^2}{Pr} \nonumber \\
	&=& \frac{\alpha g_0 \Delta T}{\Omega^2 D},
	\label{eqn:Rastar}
\end{eqnarray}
and a modified flux-based Rayleigh number,
\begin{eqnarray}
	Ra_Q^* &=& \frac{1}{4\pi r_o r_i} \frac{\alpha g_0 Q_{adv}}{\rho c \Omega^3 D^2} \nonumber \\
	&=& Ra^* Nu^* \label{eqn:RaQstar}\\
	&=& (Nu-1) \frac{Ra \hspace{2pt} Ek^3}{Pr^2}, \nonumber
\end{eqnarray}
are used, neither of them containing any diffusivity. On the basis of these diffusivity-free parameters, \citet{Christensen:2006p11} studied the scaling of heat transport, flow velocity and magnetic field strength in numerical dynamo models. The preferred scalings for all three quantities were simple power laws only depending on $Ra_Q^*$. In this section, we use our model selection procedure by LOOCV in order to study whether a data-driven analysis yields the same result as the diffusivity-free hypothesis.

\subsection{Heat transport}
\label{sec:diffless_heat_transport}

\begin{figure*}
\centering
\includegraphics[width=\columnwidth]{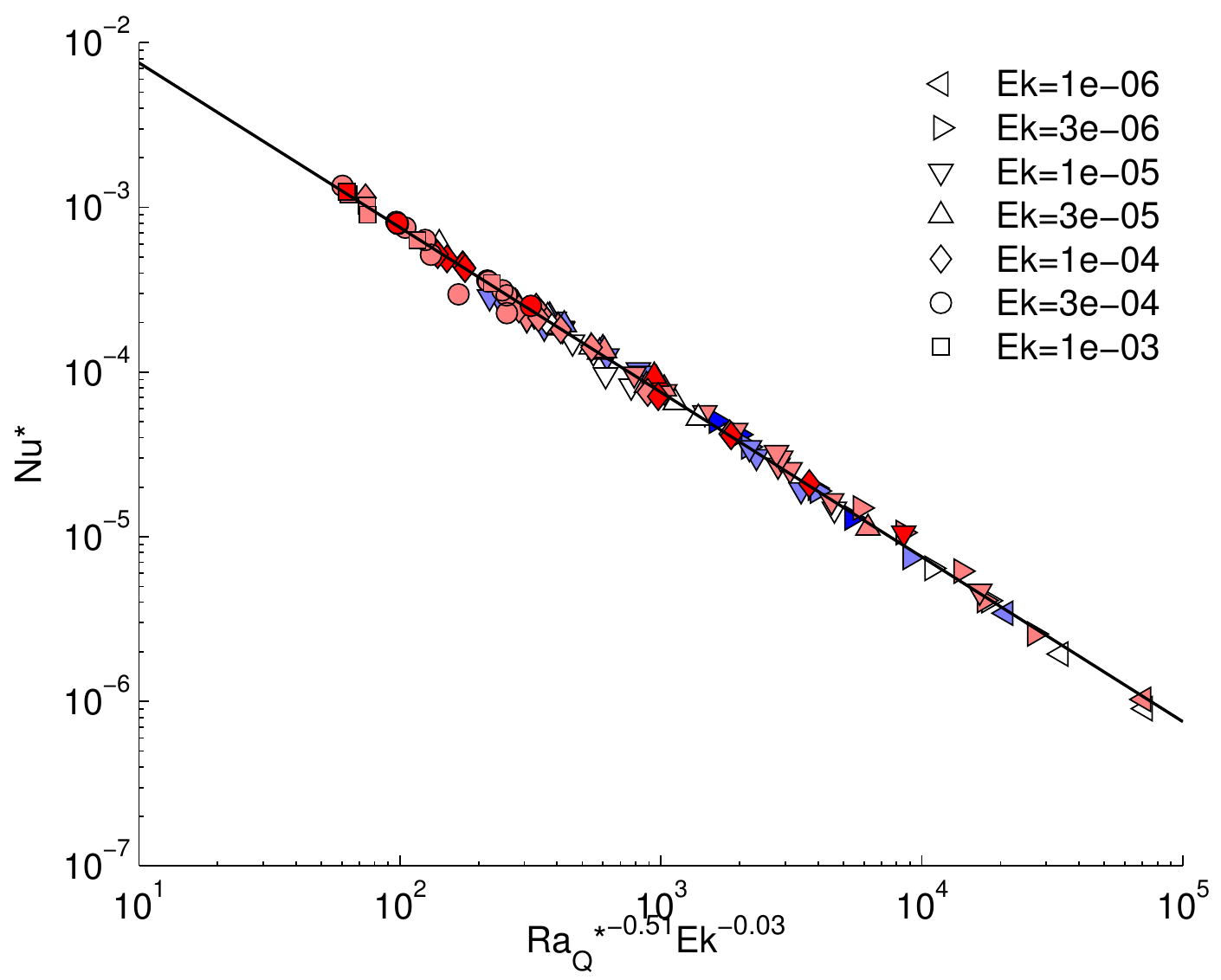}
\caption{Heat transport scaling, preferred scaling law by LOOCV. Colour indicates the value of $Pm$: dark-blue $Pm\le0.1$, light-blue $0.1<Pm<1$, white $Pm=1$, light-red $1<Pm<10$, dark-red $Pm\ge10$.}
\label{fig:C116_Nustar_RaQstar_Ek}
\end{figure*}

\begin{table*}
\begin{minipage}{0.9\textwidth}
\caption{Cross-validation estimates of prediction error $P_{CV}$ for the best-fitting scaling laws for heat transport, flow velocity and magnetic field strength for all possible parameter combinations. Minimum values are again bold.}
\centering
\begin{tabular}{|c|c c c c c c c|}
\hline
			& $(Ra_Q^*)$ 	& $(Pm)$	& $(Ek)$	& $(Ra_Q^*, Pm)$	& $(Ra_Q^*, Ek)$	& $(Pm, Ek)$	& $(Ra_Q^*, Pm, Ek)$ \\ \hline
$Nu^*$		& 0.0106		& 2.5772	& 1.0396	& 0.0100			& \bf{0.0095}		& 0.8412		& 0.0096				\\
$Ro$		& 0.0438		& 1.8391	& 0.9118	& \bf{0.0116}		& 0.0315			& 0.6164		& 0.0118				\\
$Lo/f_{ohm}^{1/2}$ & 0.0760	& 0.9238	& 0.3486	& \bf{0.0264}		& 0.0580			& 0.3466		& 0.0266 				\\ \hline
\end{tabular}
\label{tab:Nustar_scaling_PCV}
\end{minipage}
\end{table*}%
The heat transport in terms of diffusivity-free parameters is given by $Nu^*$. We test scaling laws of the form of equation \ref{eqn:exp_scal_form} and allow any combination of $Ra_Q^*$, $Pm$ and $Ek$ as explanatory variables. The cross-validation estimates of the prediction error $P_{CV}$ for the best-fitting laws with all different parameter combinations are given in Table \ref{tab:Nustar_scaling_PCV}. The scaling law with minimum $P_{CV}$ includes the parameters $Ra_Q^*$ and $Ek$:
\begin{equation}
	\label{eqn:Nustar_scaling}
	Nu^* = 0.075 \hspace{1mm} Ra_Q^{*0.51} Ek^{0.03}.
\end{equation}
Comparably low $P_{CV}$ result from scaling laws including the parameter combinations $(Ra_Q^*)$, $(Ra_Q^*, Pm)$ and $(Ra_Q^*, Pm, Ek)$. Table \ref{tab:diffless_scalings} shows the fitted values of equation \ref{eqn:Nustar_scaling} together with their standard errors. The table also contains the mean relative misfit $\chi_{rel}$ defined in equation \ref{eqn:mean_rel_misfit}.\\
Figure \ref{fig:C116_Nustar_RaQstar_Ek} shows the fit of the scaling law (eq. \ref{eqn:Nustar_scaling}) to the 116 data points. Disregarding the additional $Ek$-dependence, the scaling is very similar to $Nu^* = 0.076 \hspace{1mm} Ra_Q^{*0.53}$ \citep{Christensen:2006p11}. Although the exponent of $Ek$ is quite small, LOOCV, under the assumption of equal errors in $\zeta=\log(Nu^*)$, argues for this dependence, and the numerical value of the exponent is four times larger than its standard error in regression. One reason for the weak $Ek$-dependence could be that an asymptotic behavior has not yet been reached within the rapidly-rotating regime \citep[cf.][]{King:2010p5}. Also, it should be mentioned that LOOCV under the assumption of equal errors in the original variable $y=Nu^*$ favours a simple $Ra_Q^*$-dependence devoid of any $Ek$-dependence (see appendix \ref{sec:equal_errors_y}).
\\
We compare our heat flux scaling relation with others that have recently appeared in the literature. \citet{King:2010p5} compare scaling
relations developed for experiments in rotating cylinders (in which gravity is parallel to the rotation axis) with the same type of
numerical results that are analysed herein, namely rotating convection with radial gravity. For the rapidly-rotating regime, they find a preferred fit to their experimental data that is also in reasonably good agreement with the numerical results of the form
\begin{equation}
	Nu=A\left(\frac{Ra}{Ra_c}\right)^{6/5},
\end{equation}
where $Ra_c\propto Ek^{-4/3}$ is the critical Rayleigh number for the onset of convection. In terms of the flux-based quantities that we are considering here, this law becomes
\begin{equation}
	Nu^* \propto (Ra_Q^*)^{6/11}(Ek \hspace{0.5mm} Pr)^{1/11} .
\end{equation}
The numerical values of these indices, 0.545 and 0.09, are not terribly different from the ones that we discovered here.\\
Conversely, a recent explanation of the same experimental data by \citet{KingEtAl:2012} proposes
\begin{equation}
	Nu = A\left(\frac{Ra}{Ra_c}\right)^3 \propto Ra^3Ek^4
\end{equation}
based on a physically-motivated boundary layer analysis. In terms of the flux-based parameters, this is equivalent to
\begin{equation}
	Nu^* \propto (Ra_Q^*)^{3/4}Pr^{1/2}Ek^{-1/4}.
\end{equation}
The Ekman dependence of this law is clearly much stronger than others that have been proposed (including our own), and has an opposite sign of exponent when converted to flux-based variables. The lack of experimental data in the strongly rotation-dominated regime contributes to this lack of understanding.\\
\begin{table*}
\begin{minipage}{0.9\textwidth}
\caption{Overview of the scaling laws preferred by LOOCV for the diffusivity-free parameters. The exponents of the non-dimensional parameters are shown together with their standard errors from the multiple linear regression. Covariances between the fitted values are minor. The mean relative misfit $\chi_{rel}$ of the different models is also displayed.}
\centering
\begin{tabular}{|c|c c c c|c|}
\hline
			& prefactor			& $Ra_Q^*$ 			& $Pm$				& $Ek$ 			& $\chi_{rel}$ \\ \hline
$Nu^*$		& $0.075 \pm 0.004$	& $0.505 \pm 0.005$	& - 				& $0.033 \pm 0.008$ 	& 0.100 \\
$Ro$		& $1.16 \pm 0.05$		& $0.436 \pm 0.003$	& $-0.126 \pm 0.007$	& - 				& 0.106 \\ 
$Lo/f_{ohm}^{1/2}$	& $0.60 \pm 0.04$	& $0.306 \pm 0.005$	& $0.157 \pm 0.011$	& - 				& 0.161 \\ \hline
\end{tabular}
\label{tab:diffless_scalings}
\end{minipage}
\end{table*}%

\subsection{Flow velocity}

\begin{figure*}
\centering
\includegraphics[width=\columnwidth]{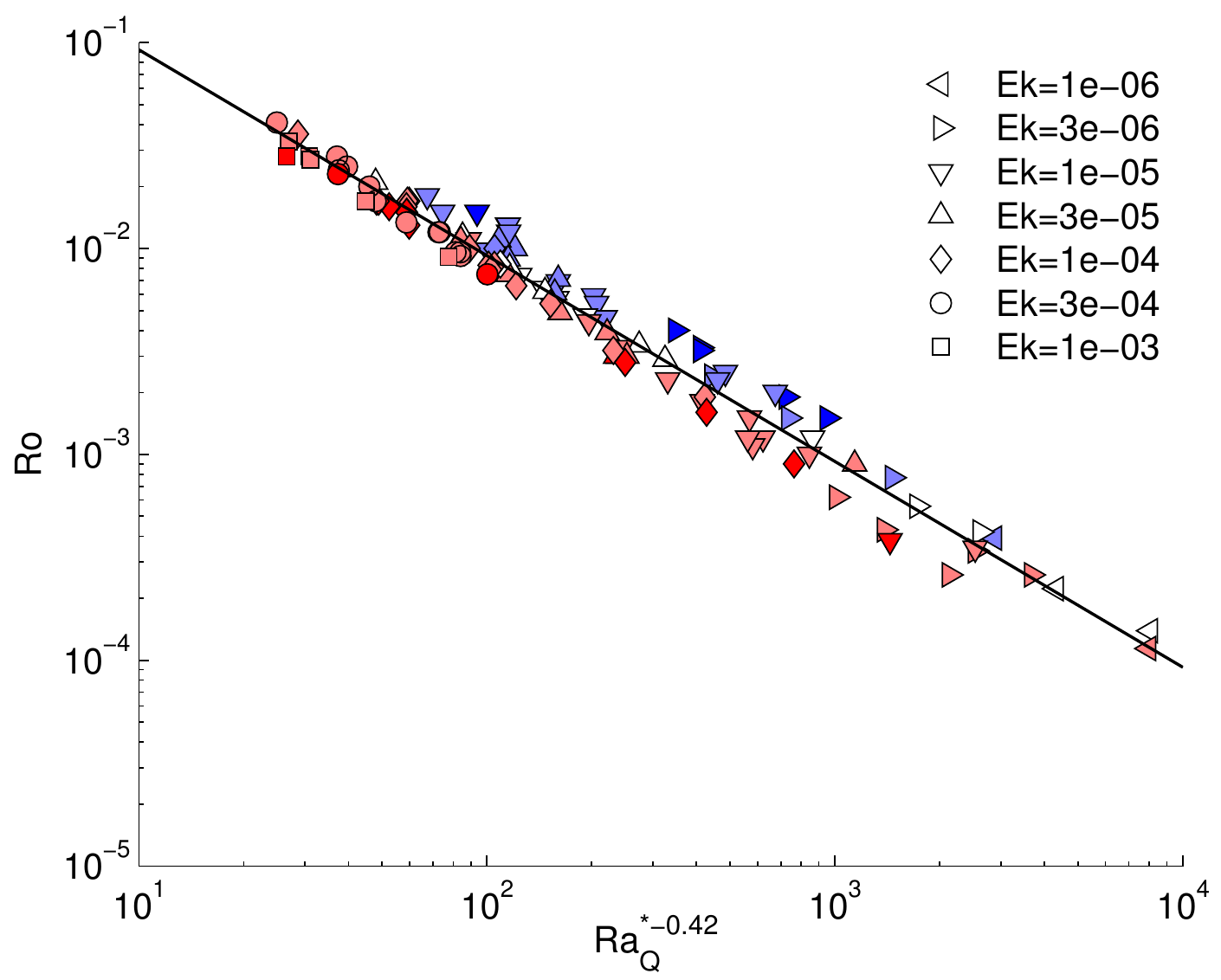}
\caption[]{
Flow velocity scaling only with $Ra_Q^*$ (not preferred by LOOCV). Colours as in Figure \ref{fig:C116_Nustar_RaQstar_Ek}. There is a clear division between blue ($Pm<1$) and red ($Pm>1$) above and below the fitting line.
}
\label{fig:Ro_scaling_RaQstar}
\end{figure*}

\begin{figure*}
\centering
\includegraphics[width=\columnwidth]{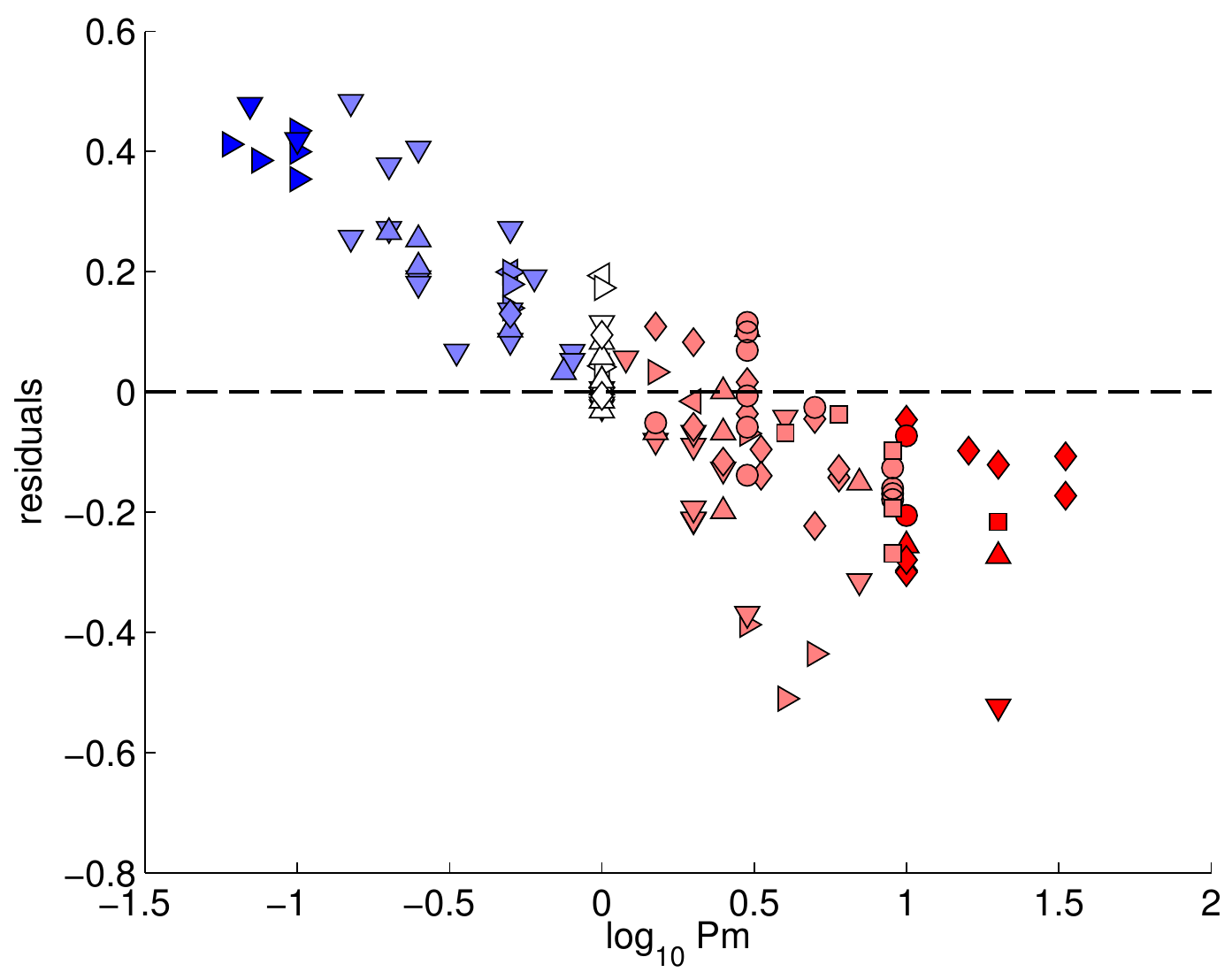}
\caption[]{
Residuals between $Ro$-data and model predictions from fig. \ref{fig:Ro_scaling_RaQstar} plotted vs. $Pm$. A clear unresolved $Pm$-dependence is visible. Colours as in Figure \ref{fig:C116_Nustar_RaQstar_Ek}.
}
\label{fig:Ro_scaling_RaQstar_resid}
\end{figure*}

A measure for flow velocity in non-dimensional form is $Ro$ as defined by 
\begin{equation}
	Ro = \left( \frac{2 E_{kin}}{V} \right) ^{1/2},
	\label{eqn:Ro}
\end{equation}
where $E_{kin}$ is kinetic energy and $V$ is the volume of the shell \citep{Christensen:2006p11}. Applying the same procedure as in section \ref{sec:diffless_heat_transport} leads to a flow velocity scaling of
\begin{equation}
	Ro = 1.16 \hspace{1mm} Ra_Q^{*0.44}Pm^{-0.13}.
	\label{eqn:Ro_scaling}
\end{equation}
\begin{figure*}
\centering
\includegraphics[width=\columnwidth]{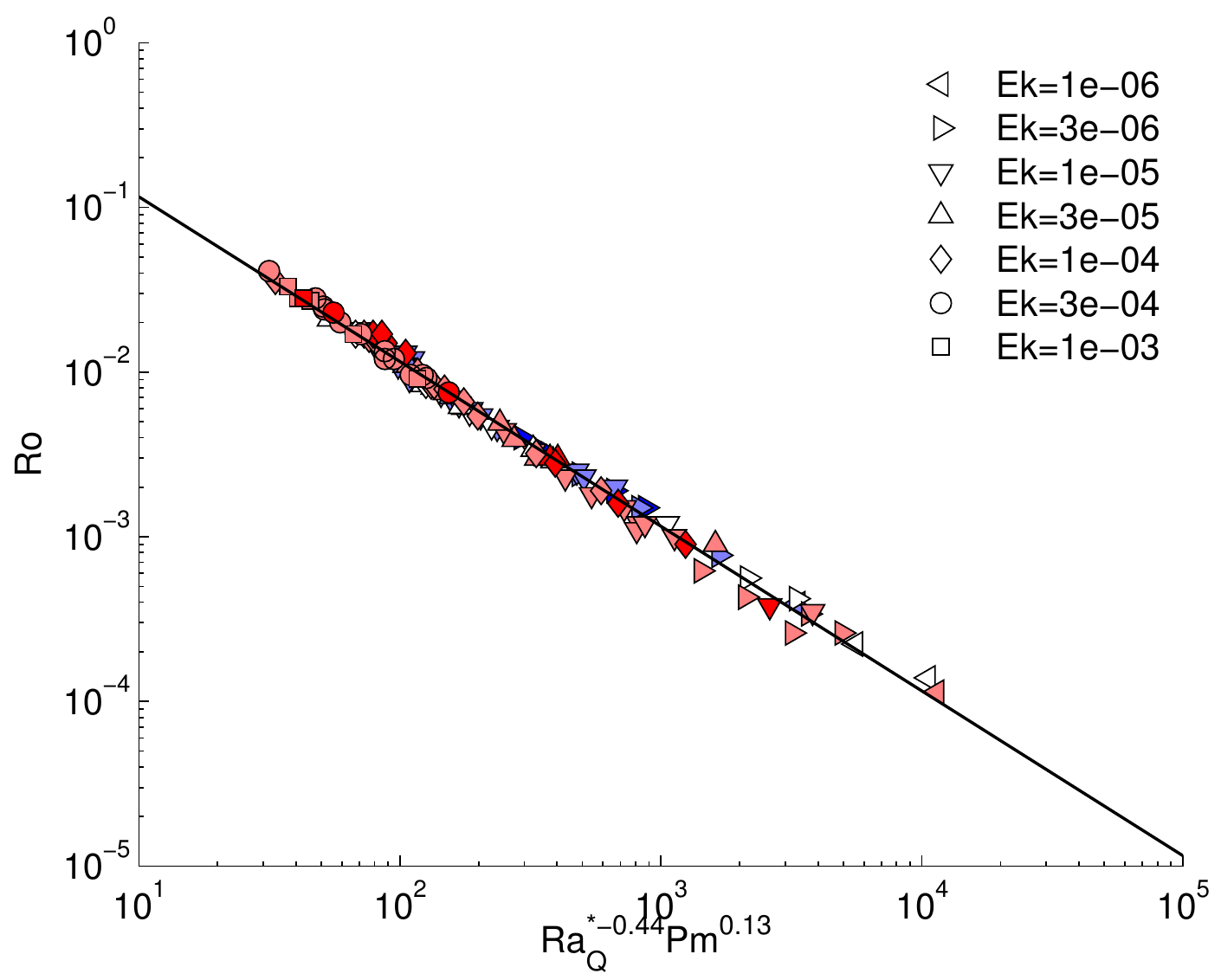}
\caption[]{
Flow velocity, favoured scaling law.  Colours as in Figure \ref{fig:C116_Nustar_RaQstar_Ek}.
}
\label{fig:Ro_scaling_RaQstar_Pm}
\end{figure*}%
This scaling law is shown in Figure \ref{fig:Ro_scaling_RaQstar_Pm}. It is virtually identical to the $Ra_Q^{*0.43}Pm^{-0.13}$ law that could not firmly be established by \citet{Christensen:2006p11} because the improvement in misfit compared to the one-parameter law $Ro = 0.85 \hspace{1mm} Ra_Q^{*0.41}$ did not seem to be sufficient. According to our analysis, however, $Pm$ plays a role in the $Ro$-scaling with $P_{CV}(Ra_Q^*, Pm)=0.0116$ compared to $P_{CV}(Ra_Q^*)=0.0438$ arguing for the additional dependence (cf. Table \ref{tab:Nustar_scaling_PCV}). This becomes also evident in Figure \ref{fig:Ro_scaling_RaQstar} where the one-parameter fit (including only $Ra_Q^*$) to the velocity data is shown, and in Figure \ref{fig:Ro_scaling_RaQstar_resid} where the corresponding residuals are plotted versus $Pm$. An unresolved $Pm$-dependence is visible.

\subsection{Magnetic field strength}

\begin{figure*}
\centering
\includegraphics[width=\columnwidth]{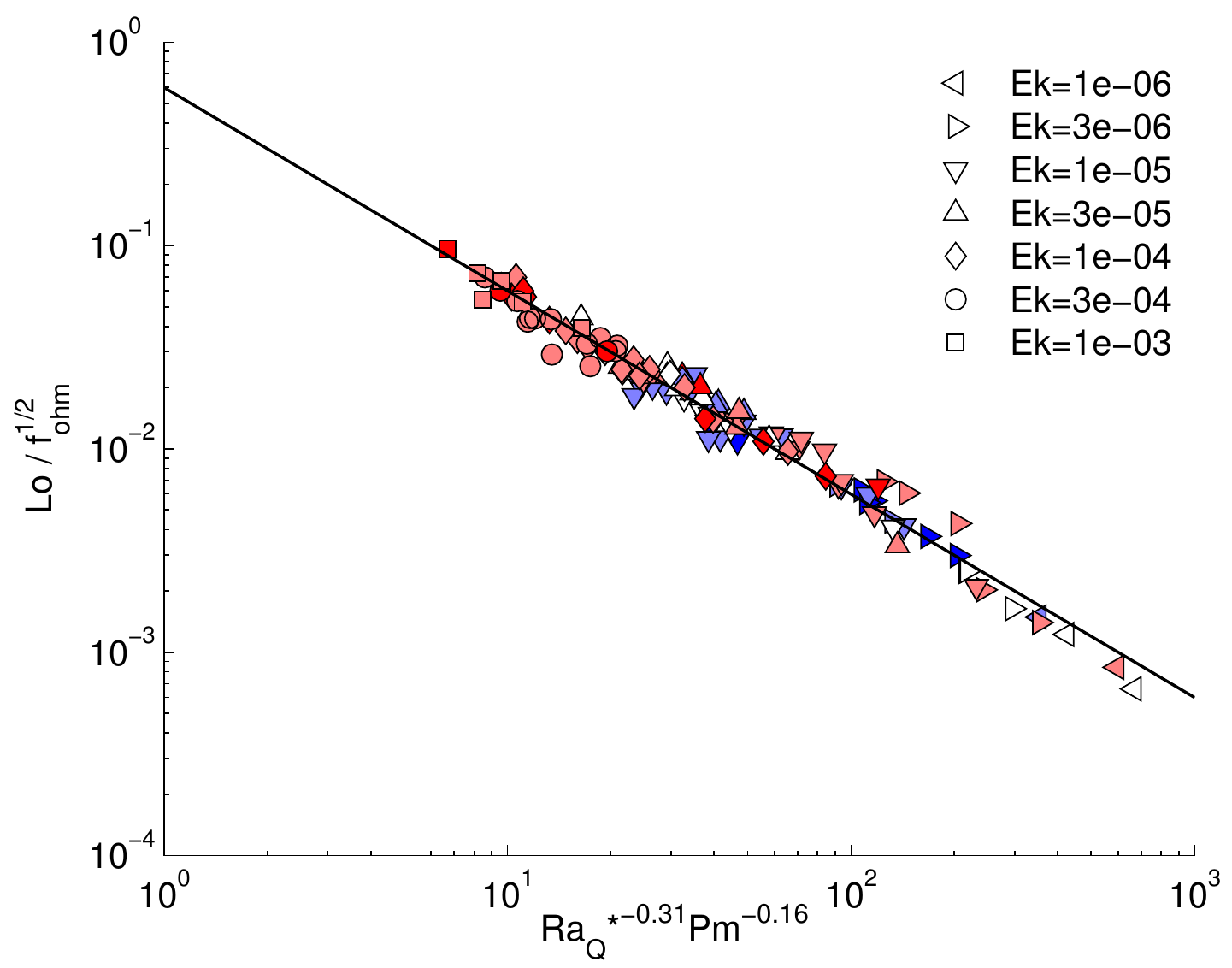}
\caption[]{
Magnetic field strength, favoured scaling law.  Colours as in Figure \ref{fig:C116_Nustar_RaQstar_Ek}.
}
\label{fig:Lofohm_scaling_RaQstar_Pm}
\end{figure*}

An adequate measure for magnetic field strength is given by $Lo/f_{ohm}^{1/2}$ according to \citet{Christensen:2006p11}. The Lorentz number $Lo$ is defined analogously to $Ro$ (eq. \ref{eqn:Ro}) as
\begin{equation}
	Lo = \left( \frac{2 E_{mag}}{V} \right) ^{1/2},
	\label{eqn:Lo_def}
\end{equation}
with magnetic energy replacing kinetic energy. The time-averaged fraction of Ohmic dissipation,
\begin{equation}
	f_{ohm} = \frac{D_{ohm}}{P},
	\label{eqn:fohm_def}
\end{equation}
is the ratio of Ohmic dissipation,
\begin{eqnarray}
	D_{ohm} &=& \int \textbf{j}^2 / \sigma \hspace{2pt} dV \nonumber \\
	&=& \int (\eta / \mu_0) (\nabla \times \textbf{B})^2 \hspace{2pt} dV,
	\label{eqn:ohmic_dissipation}
\end{eqnarray}
to the power $P$ generated by buoyancy forces; $\textbf{j}$ is the electrical current density.\\
Again, we look for a scaling of power law form that includes any combination of $Ra_Q^*$, $Pm$ and $Ek$. The law favoured by our model selection analysis is
\begin{equation}
	\frac{Lo}{f_{ohm}^{1/2}} = 0.60 \hspace{1mm} Ra_Q^{*0.31} Pm^{0.16}.
	\label{eqn:Lofohm_scaling}
\end{equation}
It is shown in Figure \ref{fig:Lofohm_scaling_RaQstar_Pm}. Also in this case, our analysis differs from \citet{Christensen:2006p11} who preferred the one-parameter scaling $Lo/f_{ohm}^{1/2} = 0.92 \hspace{1mm} Ra_Q^{*0.34}$ over the $Ra_Q^{*0.32} Pm^{0.11}$ law. (The exponent of $Pm$ in a two-parameter law for $Lo/f_{ohm}^{1/2}$ has risen from 0.11 in the original study to 0.16 in eq. \ref{eqn:Lofohm_scaling}, probably due to adding dynamo models with large $Pm$ to the dataset.) The estimated prediction errors are $P_{CV}(Ra_Q^*, Pm)=0.0264$ versus $P_{CV}(Ra_Q^*)=0.0760$ favouring the additional dependence (cf. Table \ref{tab:Nustar_scaling_PCV}).

\subsection{Discussion}

The diffusivity-free scalings for heat transport, flow velocity and magnetic field strength contain only a dependence on $Ra_Q^*$ \citep{Christensen:2006p11}. Our model selection analysis by LOOCV, however, favours more complex scalings with an additional parameter. As mentioned in section \ref{sec:diffless_heat_transport}, the $Ek$-dependence in the $Nu^*$-scaling may be due to a non-asymptotical regime and disappears when a different error attribution is used (Appendix \ref{sec:equal_errors_y}). The $Pm$-dependence in the scalings of $Ro$ and $Lo/f_{ohm}^{1/2}$ is a significant feature which also persists when using different methods of model selection.\\
Summing up, we used diffusivity-free parameters in the first place. But the diffusivities come back into the scaling laws by additional dependencies complicating the simple laws. This means that diffusive processes may not be neglected in the regime of numerical dynamo models that we are looking at (cf. section \ref{sec:dynamical_regime}) An attempt to apply the scaling laws to Earth's core is undertaken in section \ref{sec:implications}.


\section{Scaling with traditional parameters}
\label{sec:trad_scal}

In the previous section, we have shown that the numerical dynamo simulations in general do not support diffusivity-free scalings of heat transport, flow velocity and magnetic field strength. The question is now about the scalings in terms of traditional parameters $Ra$, $Pm$, $Ek$ and $Pr$ (definitions in Table \ref{tab:nondim_no}). In this case, it is necessary to allow a possible $Pr$-dependence in order to account for the variability in the data. (The diffusivity-free parameter $Ra_Q^*$ in equation \ref{eqn:RaQstar} has an implicit $Pr$-dependence.)\\
Again, we look for exponential scaling laws of the form of equation \ref{eqn:exp_scal_form}. LOOCV favours the following scaling laws for convective heat transport, flow velocity and magnetic field strength, respectively:
\begin{eqnarray}
	Nu-1 &=& 0.009 \hspace{1mm} Ra^{0.93} Ek^{1.00} Pr^{-0.09}, \label{eqn:Nu_trad_scaling}
\end{eqnarray}
\begin{eqnarray}
	Ro &=& 0.15 \hspace{1mm} Ra^{0.84} Pm^{-0.13} Ek^{1.75} Pr^{-0.90}, \label{eqn:Ro_trad_scaling} 
\end{eqnarray}
\begin{eqnarray}
	\frac{Lo}{f_{ohm}^{1/2}} &=& 0.18 \hspace{1mm} Ra^{0.54} Pm^{0.17} Ek^{1.15} Pr^{-0.71}. \label{eqn:Lofohm_trad_scaling}
\end{eqnarray}
(We choose $(Nu-1)$ as measure of convective heat transport and stay with $Lo/f_{ohm}^{1/2}$ as measure of magnetic field strength in order to get laws that are comparable with the scalings of section \ref{sec:diff_free_scal}. In the case of the magnetic field scaling, it should be noted that even the simplest law, $Lo/f_{ohm}^{1/2} \sim Ra_Q^{*\beta}$, is actually not diffusivity-free in general, since $f_{ohm}$ contains the magnetic diffusivity $\eta$ via the Ohmic dissipation $D_{ohm}$, cf. eqs. \ref{eqn:fohm_def} and \ref{eqn:ohmic_dissipation}. The scaling is only diffusivity-free when $f_{ohm} \approx 1$ as assumed for Earth's core.)\\
The dependencies in equations \ref{eqn:Nu_trad_scaling}-\ref{eqn:Lofohm_trad_scaling} are complex enough to require all parameters in the scaling laws. Only in the $(Nu-1)$-scaling, $Pm$ is not included as it is the case in section \ref{sec:diffless_heat_transport}. It is, however, clear that these scalings are pure linear regression results on the data lacking any physical rationale.
\begin{table*}
\begin{minipage}{0.9\textwidth}
\caption{Overview of the scaling laws preferred by LOOCV for the traditional parameters. The exponents of the non-dimensional parameters are shown together with their standard errors from the multiple linear regression. $\chi_{rel}$ is the mean relative misfit between fitted and observed values (eq. \ref{eqn:mean_rel_misfit}).}
\centering
\begin{tabular}{|c|c c c c c|c|}
\hline
			& prefactor			& $Ra$ 			& $Pm$			& $Ek$ 			& $Pr$ 			& $\chi_{rel}$ \\ \hline
$Nu-1$		& $0.009 \pm 0.001$	& $0.93 \pm 0.02$	& - 				& $1.00 \pm 0.02$	& $-0.09 \pm 0.02$ 	& 0.165 \\
$Ro$		& $0.15 \pm 0.02$		& $0.84 \pm 0.01$	& $-0.13 \pm 0.01$	& $1.75 \pm 0.02$	& $-0.90 \pm 0.02$ 	& 0.100 \\
$Lo/f_{ohm}^{1/2}$	& $0.18\pm0.03$	& $0.54 \pm 0.02$	& $0.17 \pm 0.02$	& $1.15 \pm 0.03$	& $-0.71 \pm 0.02$ 	& 0.173 \\ \hline
\end{tabular}
\label{tab:traditional_scalings}
\end{minipage}
\end{table*}%
Table \ref{tab:traditional_scalings} shows that the scalings are quite complex. Creating diffusivity-less parameters (eq. \ref{eqn:Nustar}-\ref{eqn:RaQstar}) with inbuilt $Ek$- and $Pr$-dependencies has been an attempt to simplify the relations.\\
We can actually find a parameter similar to the modified Rayleigh number, $Ra^* = RaEk^2Pr^{-1}$ (eq. \ref{eqn:Rastar}), in the scalings for flow velocity and magnetic field strength (eq. \ref{eqn:Ro_trad_scaling} and \ref{eqn:Lofohm_trad_scaling}), when we look at the exponents of $Ra$, $Ek$ and $Pr$ that form a ratio of approximately 1 : 2 : -1 in the $Ro$- and $Lo/f_{ohm}^{1/2}$-scalings. This parameter combination is also known as the convective Rossby number, $Ro_c = (Ra^*)^{1/2}$ \citep[e.g.][]{Liu:1997p1694,Aurnou2007110}. The convective Rossby number describes the ratio of buoyancy over Coriolis forces when using the convective free-fall velocity, $u_{conv} \sim \sqrt{\alpha g_o \Delta T D}$, which results from a balance between inertia and buoyancy, as velocity scale. Hence is not surprising to find $Ra^*$ in the velocity scaling. It is slightly more surprising to see it in the magnetic field scaling, although induction scales with the velocity field. On top of the $Ra^*$-dependence, there is certainly a $Pm$-dependence present in both scalings. The heat transport scaling (eq. \ref{eqn:Nu_trad_scaling}), however, is not at all reminiscent of $Ra^*$ and does not contain a $Pm$-dependence either.


\section{Magnetic dissipation in Earth's core}
\label{sec:mag_dissipation_core}

\subsection{Magnetic dissipation time}

The magnetic dissipation time $\tau_{diss}$ is defined as the ratio of magnetic energy over Ohmic dissipation (eq. \ref{eqn:ohmic_dissipation}),
\begin{equation}
	\tau_{diss} = \frac{E_{mag}}{D_{ohm}}.
	\label{eqn:taudiss}
\end{equation}
With knowledge about $\tau_{diss}$ and an estimate of $E_{mag}$, we are able to put numbers on the Ohmic dissipation $D_{ohm}$ in Earth's core.\\
\citet{Christensen:2004p706} found an inverse dependence of $\tau_{diss}$ on the magnetic Reynolds number $Rm$. The same study rejects an additional dependence on $Re=Rm/Pm$ (which is equivalent to an additional dependence on $Pm$) because of results of the Karlsruhe laboratory dynamo. Later, \citet{Christensen:2010p1} revisited the $\tau_{diss}$-scaling favouring an additional dependence on the magnetic Ekman number $Ek_\eta = Ek/Pm$. \\
Using the magnetic diffusion time $\tau_\eta = D^2/\eta$ to normalise the magnetic dissipation time,
\begin{equation}
\label{eqn:taudissstar}
	\tau_{diss}^* = \frac{\tau_{diss}}{\tau_\eta},
\end{equation}
the 2004 and the 2010 laws are given as
\begin{eqnarray}
	\tau_{diss, 04}^* &=& 0.27 \hspace{1mm} Rm^{-1}, \label{eqn:taudiss_scaling_2004} \\
	\tau_{diss, 10}^* &=& 0.59 \hspace{1mm} Rm^{-5/6} Ek_\eta^{1/6} \nonumber \\
	&=& 0.59 \hspace{1mm} Rm^{-5/6} Pm^{-1/6} Ek^{1/6}. \label{eqn:taudiss_scaling_2010}
\end{eqnarray}

\subsection{LOOCV analysis for $\tau_{diss}$}

\begin{figure*}
\centering
\includegraphics[width=\columnwidth]{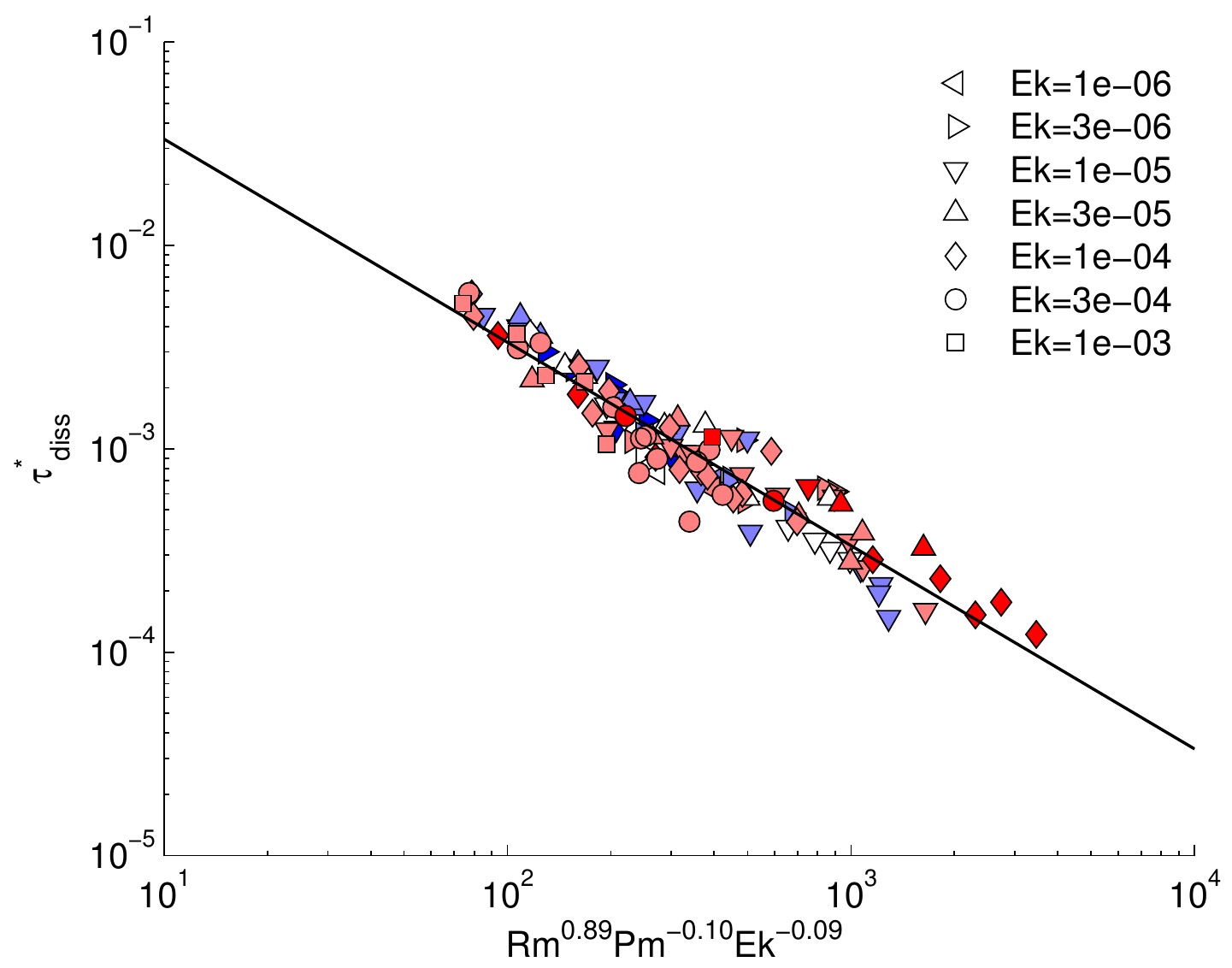}
\caption[]{
Magnetic dissipation time. Favoured scaling law. Colours as in Figure \ref{fig:C116_Nustar_RaQstar_Ek}.
}
\label{fig:taudiss_scaling_Rm_Pm_Ek}
\end{figure*}

According to the scaling laws of equations \ref{eqn:taudiss_scaling_2004} and \ref{eqn:taudiss_scaling_2010}, it seems reasonable to test scaling laws for $\tau_{diss}^*$ that have power law form including the parameters $Rm$, $Pm$, $Ek$ (and possibly $Pr$). Our model selection analysis by LOOCV on the basis of the 116 numerical dynamo models favours the full model,
\begin{equation}
	\tau_{diss}^* = 0.33 \hspace{1mm} Rm^{-0.89} Pm^{0.10} Ek^{0.09},
	\label{eqn:taudiss_scaling}
\end{equation}
shown in Figure \ref{fig:taudiss_scaling_Rm_Pm_Ek}. $P_{CV}(Rm, Pm, Ek)=0.0777$ compared to $P_{CV}(Rm)=0.1400$ and $P_{CV}(Rm, Ek_\eta)=0.1321$. The standard errors on the prefactor and on the exponents in equation \ref{eqn:taudiss_scaling} are 0.08, 0.03, 0.03 and 0.02, respectively. The mean relative misfit $\chi_{rel}$ of this scaling law is 0.289, significantly larger than for the previous scalings. (Allowing a $Pr$-dependence in the model selection procedure again leads to the full model including $Pr$ and reduces the mean relative misfit to 0.205. However, see the remarks on the distribution of $Pr$ in our dataset in section \ref{sec:par_range}.)
\begin{figure*}
\centering
\includegraphics[width=\columnwidth]{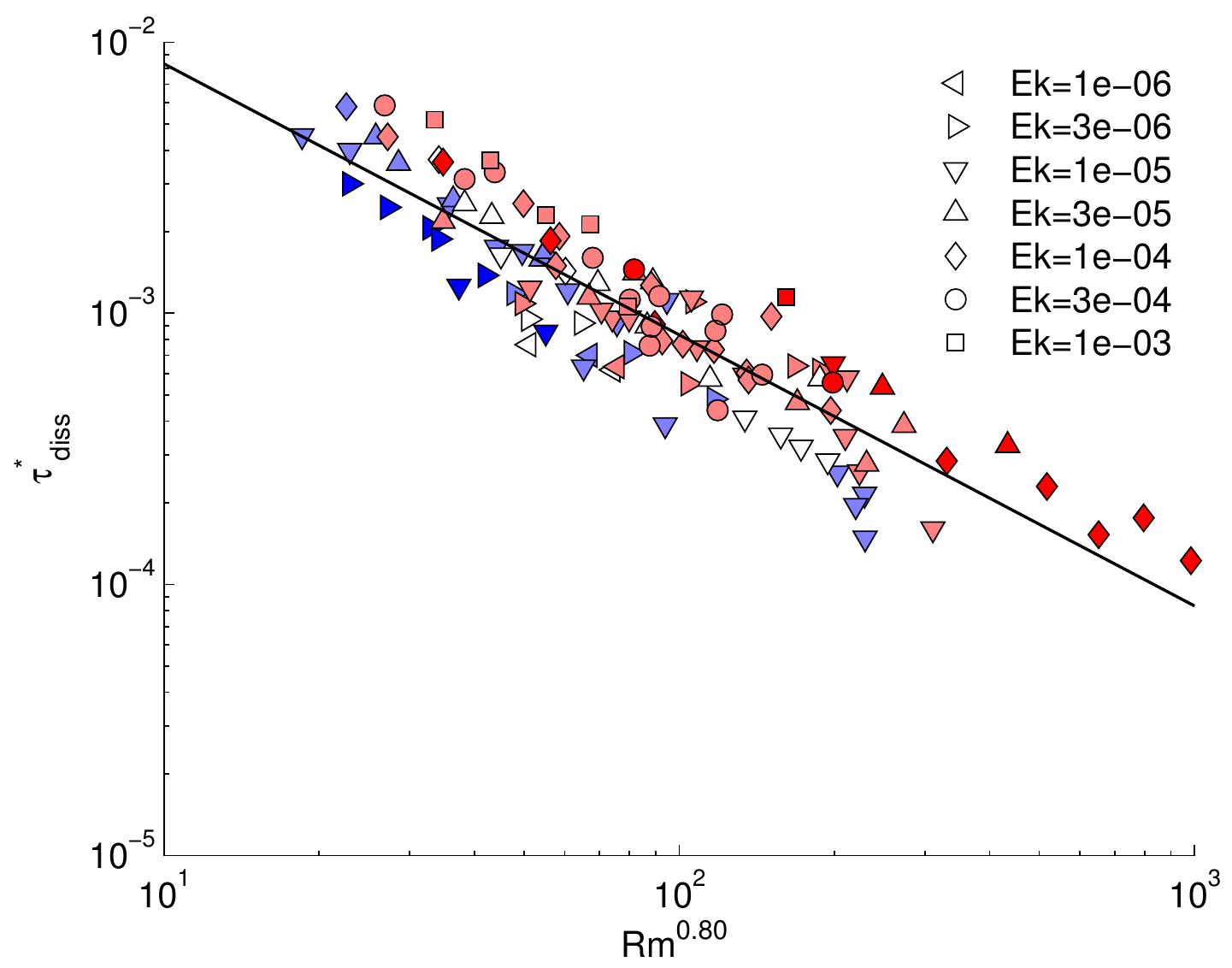}
\caption[]{
Magnetic dissipation time. Simple $Rm$-law with unresolved further dependencies. Colours as in Figure \ref{fig:C116_Nustar_RaQstar_Ek}.
}
\label{fig:taudiss_scaling_Rm}
\end{figure*}
\\A scaling law with only $Rm$ as independent variable on the basis of the 116 dynamo models would be $\tau_{diss}^* = 0.083 \hspace{1mm} Rm^{-0.80}$, displayed in Figure \ref{fig:taudiss_scaling_Rm}. While the numerical dynamo database of Christensen and co-workers has grown over the years, the exponent of $Rm$ in a simple one-parameter law for $\tau_{diss}^*$ has decreased in absolute magnitude from -1 \citep{Christensen:2004p706} via -0.93 \citet{Christensen:2010p1} to -0.8 in this study. The first thing to notice in the plot is the clear unresolved $Pm$-dependence in the data plotted according to this law. The subsets of data with equal $Ek$ (and similar $Pm$) appear to follow slopes that are similar to -0.8 with different y-axis intercepts. This would mean that $Ek$ and $Pm$ mainly determine the prefactor in the exponential scaling law. The favoured scaling law for $\tau_{diss}^*$ (eq. \ref{eqn:taudiss_scaling}) implies that this quantity grows with increasing $Pm$. This dependence is contrary to the scaling including the magnetic Ekman number (eq. \ref{eqn:taudiss_scaling_2010}).

\subsection{Application to Earth's core}
\label{sec:application_core}

One quantity that is of interest for the study of the Earth's deep interior is the amount of Ohmic dissipation in the core. \citet{Christensen:2004p706} used their scaling law for the magnetic dissipation time (eq. \ref{eqn:taudiss_scaling_2004}) to derive 42~years for the magnetic dissipation time and an estimate of 0.2-0.5~TW for the Ohmic dissipation, which was a rather small value compared to other estimates. These calculations are based on $Rm =  800$ (note the differing parameter definition in the original paper) and $E_{mag} = (2.8-6.2) \cdot 10^{20}~\mathrm{Joules}$. \citet{Christensen:2010p1} found an Ohmic dissipation time that is five times shorter and hence a five times higher value for the Ohmic dissipation using the revised scaling law (eq. \ref{eqn:taudiss_scaling_2010}).\\
We base our calculations on the current estimates for the non-dimensional parameters given in Table \ref{tab:nondim_no}. A major revision of these numbers has resulted from studies by \citet{deKoker:2012p2910} and \citet{Pozzo:2012p2852} that have increased the numerical values of the thermal and the electrical conductivities, $\kappa$ and $\sigma$, for Earth's core by roughly a factor of three. Together with flow velocities of $\sim 15~\mathrm{km/year}$ inferred from secular variation studies \citep{Bloxham:1991p2791,Holme:2007p2955}, this yields $Rm \approx 2300$. Note, however, that large uncertainties are associated with this estimate that is based on the large-scale flow only.\\
The value for the magnetic energy in \citet{Christensen:2004p706} was derived from an assumed magnetic field strength of 2-3~mT in the core that comes from considerations about the field strength at the core-mantle boundary (CMB). More recent studies of the magnetic field strength in the core found similar values. \citet{Aubert:2009p707} used two end-member scenarios, high and low power, to study the evolution of heat flow in the core. The high-power model gives a present-day r.m.s. core magnetic field of 2.3~mT, whereas the low-power model leads to a magnetic field of 1.1~mT. \citet{Buffett:2010p2851} studied tidal dissipation in the Earth's core. In this context, nutation observations can be explained by a core-averaged field strength of 2.5~mT. \citet{Gillet:2010p2821} studied variations of length-of-day (LOD) in the context of torsional waves. They estimated an r.m.s. field strength of $\sim4~\mathrm{mT}$ inside the Earth's core. Concluding, the value of 2-3~mT for the r.m.s. field strength in the Earth's core still lies in the range of recent estimates, although the value could also be slightly higher. Hence, we use the same estimate of $(2.8-6.2) \cdot 10^{20}~\mathrm{J}$ for the magnetic energy as \citet{Christensen:2004p706}.\\
Finally, we have to assume that the processes in the numerical simulations are relevant to the dynamics of Earth's core in order to be able to extrapolate using scaling laws. This is by no means certain. But we may try since Earth's core appears to reside in the rapidly-rotating regime \citep{King:2010p5} as do the numerical dynamo models of this study (cf. section \ref{sec:dynamical_regime}).\\
Under these assumptions, the scaling in equation \ref{eqn:taudiss_scaling} yields a magnetic dissipation time of 2.3~years. Using equations \ref{eqn:taudiss} and \ref{eqn:taudissstar}, this leads to an Ohmic dissipation of 3.4-8.4~TW in Earth's core. (Using a $\tau_{diss}^*$-scaling that additionally includes $Pr$ leads to a slightly higher Ohmic dissipation.) If we include the uncertainties of the non-dimensional parameters $Rm$, $Pm$ and $Ek$, the error bars will increase further. Due to the size of the exponents, however, a change in the value of $Rm$ would alter the result most as would a change in the estimate of $E_{mag}$.\\
The Ohmic dissipation contributes to the total heat flux at the CMB. For the conductive heat flux at the top of the core, \citet{deKoker:2012p2910} find 14-20~TW using their new estimate for the thermal conductivity. Also \citet{Pozzo:2012p2852} suggest high adiabatic heat flux at the CMB with 15-16~TW on the basis of the increased thermal conductivity. These estimates are higher than the 5-15~TW found from independent considerations of core temperature, geodynamo energetics and buoyancy flux of lower-mantle thermal plumes \citep{Lay:2008p2107}, which at that time were already large compared to the previously estimated 3-4~TW. Since the dissipation should be a fraction of the total heat flux through the system, the lower range of the values 3.4-8.4~TW for Ohmic dissipation in Earth's core appears to be consistent with the recent high CMB heat flux scenarios.

\subsection{Implications}
\label{sec:implications}

The scaling laws for flow velocity (eq. \ref{eqn:Ro_scaling}), magnetic field strength (eq. \ref{eqn:Lofohm_scaling}) and Ohmic dissipation time (eq. \ref{eqn:taudiss_scaling}), as defined here, are not independent (U. Christensen, pers. comm.). The parameter definitions lead to $E_{mag} \sim Lo^2$ (eq. \ref{eqn:Lo_def}), $\tau_{diss} = E_{mag} / D_{ohm}$ (eq. \ref{eqn:taudiss}) and $D_{ohm} = f_{ohm} P \sim f_{ohm} Ra_Q^*$ (eq. \ref{eqn:fohm_def}); the latter scaling is not exact, but for large enough $Nu$ almost perfectly satisfied \citep[appendix A of ][]{Christensen:2006p11}. The interdependence of the three laws enables us to predict the $\tau_{diss}$-scaling from the $Ro$- and the $Lo/f_{ohm}^{1/2}$-scalings yielding $\tau_{diss}^* \sim Rm^{-0.89} Pm^{0.09} Ek^{0.11}$. The compliance with the LOOCV-preferred scaling law (eq. \ref{eqn:taudiss_scaling}) shows the internal consistency.\\
Using the scaling laws for flow velocity (eq. \ref{eqn:Ro_scaling}) and magnetic field strength (eq. \ref{eqn:Lofohm_scaling}) as well as the parameter values from Table \ref{tab:nondim_no}, we can also extrapolate these quantities from the numerical models to Earth's core. Combining equations \ref{eqn:Ro_scaling} and \ref{eqn:Lofohm_scaling} and eliminating $Ra_Q^*$ yields
\begin{equation}
	Lo/f_{ohm}^{1/2} = 0.54 \hspace{1mm} Ro^{0.70} Pm^{0.25}.
	\label{eqn:Lofohm_Ro_Pm_scaling}
\end{equation}
Two ways are viable here: either (a) we use an estimate for the velocity in the core to derive a magnetic field strength, or (b) we do the calculation vice versa. In case (a), assuming a velocity of $\sim 15~\mathrm{km/year}$ at the core surface (see section \ref{sec:application_core}) and $f_{ohm} \approx 1$ in the core as in \citet{Christensen:2006p11}, we find a magnetic field strength
\begin{eqnarray}
	B_{rms} &=& \sqrt{\Lambda \rho \mu_0 \eta \Omega} \nonumber \\
	&=& Lo \left( \rho \mu_0 \right) ^{1/2} \Omega D
\end{eqnarray}
of $\sim 0.1~\mathrm{mT}$, where $\Lambda = Lo^2 Pm Ek^{-1}$ is the Elsasser number. This number is lower compared to the estimates in section \ref{sec:application_core} by a factor 10 to 40. In case (b), using an estimate of $\sim 3~\mathrm{mT}$ for the magnetic field strength in the core, we find $\sim 5.6~\mathrm{cm/s}$ for the velocity, which is by a factor of 100 larger than the usual estimates. So using the scaling laws for $Ro$ (eq. \ref{eqn:Ro_scaling}) and $Lo/f_{ohm}^{1/2}$ (eq. \ref{eqn:Lofohm_scaling}), either (a) the magnetic field strength is too low, or (b) the velocity is too high. The $Pm$-dependence in equations \ref{eqn:Ro_scaling}, \ref{eqn:Lofohm_scaling} and hence also \ref{eqn:Lofohm_Ro_Pm_scaling} is at variance with the scalings found by \cite{Christensen:2006p11}, whose laws lead to much better agreement between magnetic field strengths and flow velocities thought to occur in the Earth.\\
It should, however, be noted that the usual velocity estimate of $\sim 15~\mathrm{km/year}$ is only valid for the large-scale motions on the surface of the core since it is derived from secular variation data. Small-scale velocities in the core's interior might well be significantly higher. Besides, the resolution of this discrepancy might be a modification of the scaling laws in the low-$Pm$ limit. In any case, the application of the scalings of flow velocity and magnetic field strength to Earth's core remains to be addressed.\\


\section{Conclusions}
\label{sec:conclusions}

Numerical dynamo simulations can complement theoretical considerations and laboratory experiments in the goal to gain insight into Earth's core. The derivation of scaling laws has been one important way. This approach, however, involves two major difficulties. The first is that we have to make sure that the numerical models are in the same dynamical regime as Earth's core. Although numerical models can produce Earth-like magnetic fields \citep[e.g.][]{Christensen:2010p2784}, this point is by no means certain. The second task is extracting scaling laws from the data that capture all relevant parameters.\\
We have studied approaches to the second task on the basis of 116 numerical dynamo models from the database of Christensen and co-workers.  Model selection deals with the question of how many independent variables have to be included in a model (scaling law) in order to account for the variability in the data, while avoiding over-fitting. Our method of choice is leave-one-out cross-validation (LOOCV). It rates models according to their predictive abilities and ideally prevents over-fitting.\\
Using LOOCV, we have studied the diffusivity-free scalings of heat transport ($Nu^*$), flow velocity ($Ro$) and magnetic field strength ($Lo/f_{ohm}^{1/2}$) proposed by \citet{Christensen:2006p11} as well as the scaling of the magnetic diffusion time \citep{Christensen:2004p706,Christensen:2010p1}. The physical rationale leading to diffusivity-free scalings is the idea that diffusive processes do not play a major role in Earth's core. However, it turns out that in velocity and magnetic field strength scaling, an additional dependence on $Pm$ is required by the numerical dynamo data (Table \ref{tab:diffless_scalings}). (The small $Ek$-dependence in the heat transport scaling disappears under a different error attribution to the data and might be blamed on the non-asymptotical regime of the data.) The additional dependencies mean that diffusivities come back into the scalings. Hence we find that diffusive processes are relevant in the numerical dynamos. \\
Similarly, \citet{Soderlund20129} find that transitions in dynamo behaviour from dipolar to multipolar are controlled by a competition of inertial and viscous forces. This means that also in this fundamental change in the systematics of present-day numerical dynamos, (viscous) diffusivity matters.\\
The relevance of diffusive processes is also apparent from our study of scalings with traditional parameters (section \ref{sec:trad_scal}). The favoured scaling laws are complex and require almost all possible parameters. Interestingly, it is possible to find something similar to a modified Rayleigh number $Ra^*$ with an additional $Pm$-dependence in the scalings for velocity and magnetic field strength. This is not at all true for the heat transport scaling.\\
The magnetic dissipation time $\tau_{diss}^*$ is a quantity relevant to the study of Earth's core since it allows us to estimate the Ohmic dissipation. However, also the preferred $\tau_{diss}^*$-scaling is more complex than suggested in previous studies. This leads to large error bars in the estimated quantities.\\
Using the $\tau_{diss}^*$-scaling and an estimate for the magnetic energy, we derived a range of 3-8~TW for the Ohmic dissipation in Earth's core. The lower range, 3-4~TW, of these values appears to be consistent with recent high CMB heat flux scenarios \citep{Lay:2008p2107,deKoker:2012p2910,Pozzo:2012p2852}. An unresolved issue is the application of velocity and magnetic field strength scaling to the core.\\


\section{Acknowledgments}
We are grateful to U. Christensen for providing the data of the numerical dynamo simulations used in this study and for providing some arguments given in section \ref{sec:implications}. We would also like to thank F. Takahashi and K.M. Soderlund for providing data, as well as J.M. Aurnou and E.M. King for discussion. Constructive reviews by U. Christensen and an anonymous reviewer have helped to improve the manuscript. Funding for ZS by the ERC grant 247303 `MFECE' is gratefully acknowledged.


\appendix

\section{Equal errors in the original variable}
\label{sec:equal_errors_y}

\begin{table*}
\begin{minipage}{0.9\textwidth}
\caption{Overview of the scaling laws preferred by LOOCV assuming equal errors in $y$. The corresponding laws assuming equal errors in $\zeta=\log(y)$ are given in Table \ref{tab:diffless_scalings}. The exponents of the non-dimensional parameters are shown together with their standard errors from the multiple linear regression. $\chi_{rel}$ is the mean relative misfit between fitted and observed values (eq. \ref{eqn:mean_rel_misfit}).}
\centering
\begin{tabular}{|c|c c c c|c|}
\hline
			& prefactor			& $Ra_Q^*$ 			& $Pm$				& $Ek$ 			& $\chi_{rel}$ \\ \hline
$Nu^*$		& $0.083 \pm 0.004$	& $0.545 \pm 0.005$	& - 					& - 				& 0.137 \\
$Ro$		& $1.20 \pm 0.07$		& $0.471 \pm 0.006$	& $-0.098 \pm 0.006$	& $-0.034\pm0.007$	& 0.123 \\ 
$Lo/f_{ohm}^{1/2}$	& $0.59 \pm 0.05$	& $0.302 \pm 0.008$	& $0.147 \pm 0.010$	& - 				& 0.174 \\ \hline
\end{tabular}
\label{tab:diffless_scalings_weighted}
\end{minipage}
\end{table*}%
In section \ref{sec:errors}, we discuss two possibilities of attributing errors to the data. Either we assume equal errors in $\zeta = \log(y)$ as above, or equal errors in the original measured variable $y$. Table \ref{tab:diffless_scalings_weighted} lists the scaling laws that are preferred by LOOCV under the second assumption when we allow the parameters $Ra_Q^*$, $Pm$ and $Ek$ to enter the laws as in section \ref{sec:diff_free_scal}.\\
There are two major differences between the scaling laws derived under the assumption of equal errors in $\zeta$ (Table \ref{tab:diffless_scalings}) and the ones with equal errors in $y$ (Table \ref{tab:diffless_scalings_weighted}). In the first case, the $Nu^*$-law exhibits an $Ek$-dependence, whereas in the second case it does not. However, in the second case, the $Ro$-law additionally depends on $Ek$. In order to check the validity of the assumption of Gaussian errors either in $\zeta$ or in $y$, we looked at the histograms of the residuals resulting from the two $Nu^*$-laws. In both cases, the assumption of Gaussian errors seems to be justified.

\section{Reduced dataset: Earth-like dynamo models}
\label{sec:earth_like_models}

\begin{table*}
\begin{minipage}{0.9\textwidth}
\caption{Earth-like dynamo models: Overview of the scaling laws preferred by LOOCV for the diffusivity-free parameters assuming equal errors in $\zeta=\log(y)$. The exponents of the non-dimensional parameters are shown together with their standard errors from the multiple linear regression. $\chi_{rel}$ is the mean relative misfit between fitted and observed values (eq. \ref{eqn:mean_rel_misfit}).}
\centering
\begin{tabular}{|c|c c c c|c|}
\hline
			& prefactor			& $Ra_Q^*$ 			& $Pm$				& $Ek$ 			& $\chi_{rel}$ \\ \hline
$Nu^*$		& $0.069 \pm 0.007$	& $0.479 \pm 0.009$	& - 					& $0.054\pm 0.013$& 0.114 \\
$Ro$		& $1.49 \pm 0.08$		& $0.460 \pm 0.004$	& $-0.126 \pm 0.008$	& -				& 0.075 \\ 
$Lo/f_{ohm}^{1/2}$	& $0.38 \pm 0.04$	& $0.268 \pm 0.008$	& $0.179 \pm 0.016$	& - 				& 0.155 \\ \hline
\end{tabular}
\label{tab:earthlike_diffless_scalings}
\end{minipage}
\end{table*}%
Only considering models that lie in the `Earth-like triangle' for magnetic field morphology in Figure \ref{fig:C116_Rm_Eketa_triangle} \citep[criteria of][]{Christensen:2010p2784}, the dynamo dataset is reduced from 116 to 61 models. Table \ref{tab:earthlike_diffless_scalings} shows the scaling laws that in this case are preferred by LOOCV under the assumption of equal errors in $\zeta=\log(y)$. Although the dataset is reduced by almost half, the resulting laws only differ in their exponents (up to $\pm 0.04$), but not in the parameters included (cf. Table \ref{tab:diffless_scalings}).



\end{multicols}

\end{document}